\let\OriginalLaTeXLabel\label
\documentclass[aps,prd,12pt,nofootinbib]{revtex4-2}
\let\label\OriginalLaTeXLabel
\usepackage[top=2.5cm,bottom=2.5cm,left=2cm,right=2cm]{geometry}
\usepackage[T1]{fontenc}
\usepackage{amsmath,amssymb,amsfonts}
\usepackage{graphicx}
\usepackage{slashed}
\usepackage{dsfont}
\usepackage{mathrsfs}
\usepackage{pifont}
\usepackage{physics}
\usepackage{multirow}
\usepackage{xcolor}
\usepackage{cancel}
\usepackage{booktabs}
\usepackage[caption=false]{subfig}
\usepackage[hyperindex=true,breaklinks=true,colorlinks=true,citecolor=blue,linkcolor=blue]{hyperref}
\numberwithin{equation}{section}

\allowdisplaybreaks
\newcommand{\eq}[1]{\begin{align}#1\end{align}}

\newcommand{\xmark}{\ding{55}}%

\begin{document}
\thispagestyle{empty}
\title{\texorpdfstring
{Lepton-flavor-violating trilepton tau decays induced by $S_1$ and $R_2$ scalar leptoquarks}
{Lepton-flavor-violating trilepton tau decays induced by S1 and R2 scalar leptoquarks}
}

\author{G. Hern\'andez-Tom\'e}
\email{gerardo_hernandez@uaeh.edu.mx}
\affiliation{Área Académica de Matemáticas y Física, Universidad Autónoma del Estado de Hidalgo,
Carretera Pachuca-Tulancingo Km. 4.5, Código Postal 42184, Pachuca, Hgo.}

\author{J. P. Hoyos-Daza}
\email{juan.hoyos@cinvestav.mx} 
\affiliation{ {\it Departamento de F\'isica, Centro de Investigaci\'on y de Estudios Avanzados del Instituto Polit\'ecnico Nacional} \\ {\it Apartado Postal 14-740, 07000 Ciudad de M\'exico, M\'exico}}

\author{O. G. Miranda}
\email{omar.miranda@cinvestav.mx} 
\affiliation{ {\it Departamento de F\'isica, Centro de Investigaci\'on y de Estudios Avanzados del Instituto Polit\'ecnico Nacional} \\ {\it Apartado Postal 14-740, 07000 Ciudad de M\'exico, M\'exico}}

\author{R. Sánchez-Vélez}
\email{rsanchezve@ipn.mx} 
\affiliation{ {\it Departamento de F\'isica, Centro de Investigaci\'on y de Estudios Avanzados del Instituto Polit\'ecnico Nacional} \\ {\it Apartado Postal 14-740, 07000 Ciudad de M\'exico, M\'exico}}

\begin{abstract}
Charged-lepton flavor violation is a sensitive probe of physics
beyond the Standard Model. In this work, we study the flavor-violating trilepton tau decays $\tau^- \to \ell_i^- \ell_j^- \ell_k^+$ ($\ell_{i,j,k}=e,\mu$) induced by the scalar leptoquarks $S_1$ and $R_2$. Particular emphasis is placed on $\tau^- \to e^- e^- \mu^+$ and $\tau^- \to \mu^- \mu^- e^+$, which receive contributions exclusively from box diagrams at one loop. We derive analytical expressions for the corresponding branching ratios. The charged-lepton loop amplitudes displayed here include the up-type-quark contributions from $S_1$ and $R_2^{5/3}$. Our numerical analysis considers benchmarks with real, positive Yukawa couplings to a single up-type quark generation and a representative TeV-scale leptoquark mass. We impose a selected set of phenomenological constraints from the muon anomalous magnetic moment, coherent $\mu$--$e$ conversion at tree level, radiative and three-body charged-lepton decays, lepton-flavor-violating $Z$ decays, and selected rare kaon decays. Although the branching ratios of the exclusively box-induced tau decays remain below projected experimental sensitivities within the benchmarks studied, these channels probe complementary combinations of leptoquark Yukawa couplings.
\end{abstract}
\maketitle

\section*{Introduction}

Among the theoretical frameworks proposed to address open
questions of the Standard Model (SM), leptoquark models have
emerged as well-motivated candidates.
Leptoquarks (LQs) are hypothetical particles that mediate
interactions between quarks and leptons and are predicted
in a variety of beyond-the-Standard-Model (BSM) scenarios,
including grand unified theories
(GUTs)~\cite{PhysRevLett.32.438,FRITZSCH1975193}
and composite models~\cite{Gripaios2010,Barbieri2016}.
In GUTs, quarks and leptons can belong to the same multiplet
of an enlarged gauge group, in contrast to the SM, where
they transform in distinct representations.
In such BSM frameworks, LQs may arise as scalar or vector
states characterized by their couplings to quarks and leptons,
thereby mediating quark--lepton transitions.

Furthermore, LQ models have been proposed as frameworks
capable of addressing several shortcomings of the SM.
For example, they can contribute to the generation of
neutrino masses~\cite{Zhang2021,Bigaran2019,PhysRevD.62.073009,Chen2025},
and they also appear in cosmological settings, where they
may play a role in dark matter
scenarios~\cite{Mandal2018,MOHAMADNEJAD2019114793,PhysRevD.91.035006}
and baryogenesis mechanisms~\cite{HATI2020114985}.
In addition, considerable interest has been devoted to LQs
because of their potential to account for flavor anomalies
in semileptonic $B$-meson
decays~\cite{PhysRevD.109.075028,Dorsner2013,Gripaios2015,PhysRevD.101.015011,Barbieri2016U2}%
\footnote{These examples are representative but not exhaustive.}.

Null results from a broad range of LHC searches place
stringent constraints on BSM
scenarios~\cite{Aad2020,PhysRevD.100.052013,Aad2020Dijet,Sirunyan2021}.
Nevertheless, low-energy phenomenology remains a powerful
probe of new physics.
In particular, heavy leptoquarks can generate effective
interactions at low energies once they are integrated out,
leaving observable imprints in precision measurements.
Among the possible realizations, scalar leptoquarks are
especially suitable for studying loop-induced effects,
since the low-energy phenomenology of vector leptoquarks
depends strongly on their ultraviolet
completion~\cite{DORSNER20161}.
Scalar leptoquarks can induce processes that are forbidden
or highly suppressed in the SM, such as charged-lepton
flavor violation (CLFV)~\cite{DORSNER20161,Crivellin2021,PhysRevD.93.015010}.

One key precision observable in BSM phenomenology is the
muon anomalous magnetic moment,
$a_\mu \equiv (g_\mu-2)/2$.
For many years, a persistent discrepancy between the
experimental measurement and SM predictions motivated
extensive studies of BSM contributions.
This tension, observed in the Brookhaven E821
measurement~\cite{PhysRevD.73.072003} and reinforced by the
first Fermilab Muon $g-2$
result~\cite{PhysRevLett.126.141801}, strongly influenced
BSM model building, including leptoquark scenarios, by
serving both as a motivation and as a stringent constraint.
The experimental measurement~\cite{7clf-sm2v} and
SM evaluation~\cite{ALIBERTI20251} adopted in this work
are compatible within uncertainties.
We emphasize that the significance of the discrepancy
depends on the theoretical determination of the hadronic
vacuum polarization contribution.
In light of this situation, a reassessment of the
corresponding BSM phenomenology is warranted.
We therefore use $a_\mu$ as a consistency constraint,
without requiring leptoquark contributions to explain
a significant nonzero discrepancy.

In this work, we revisit the phenomenology of scalar leptoquark models, focusing on charged-lepton-flavor-violating three-body tau decays induced by $S_1$ and $R_2$ at the TeV scale. We study the channels $\tau^- \to \ell_i^- \ell_j^- \ell_k^+$ ($\ell_{i,j,k}=e,\mu$), with particular emphasis on $\tau^- \to e^- e^- \mu^+$ and $\tau^- \to \mu^- \mu^- e^+$, which receive contributions exclusively from box diagrams at one loop. Box-diagram contributions have also been considered in several contexts~\cite{Crivellin2014,PhysRevD.108.075014,Iguro2022,BABU2010130}. The absence of photon- and $Z$-penguin contributions in
these two channels allows us to isolate the box contributions
and study the associated products of four Yukawa couplings.

Our numerical analysis considers separate up-only, charm-only, and top-only benchmarks, in which the Yukawa couplings in the up-type quark sector are restricted to a single quark flavor at a time. The nonzero couplings in this sector are taken to be real and positive, and the leptoquark mass is fixed to a representative value at the TeV scale.

In addition to $a_\mu$, we consider constraints from radiative decays $\ell_i^- \to \ell_j^- \gamma$ ($i\neq j$), coherent $\mu$--$e$ conversion in nuclei induced by tree-level leptoquark exchange, lepton-flavor-violating three-body charged-lepton decays, lepton-flavor-violating $Z$ decays, and selected rare kaon decays. The retained points therefore satisfy a selected set of constraints and are not claimed to be globally allowed. Within the scanned benchmarks, the branching ratios of the exclusively box-induced tau decays remain below projected experimental sensitivities. Nevertheless, these channels probe complementary combinations of leptoquark Yukawa couplings, and their study quantifies the suppression of the corresponding branching ratios under the imposed constraints.

The remainder of this paper is organized as follows.
In Sec.~\ref{model}, we introduce the effective Lagrangian
describing the interactions of scalar leptoquarks with
SM fermions and establish the notation used throughout
the manuscript.
In Sec.~\ref{constraints}, we discuss the low-energy
observables used to constrain the scalar leptoquark
couplings.
In Sec.~\ref{RDecay}, we present a detailed calculation
of the branching ratios for the decays
$\tau^- \to \ell_i^- \ell_i^- \ell_j^+$ ($i\neq j$)
induced by the $S_1$ and $R_2$ leptoquarks, with additional
technical details provided in the Appendices.
In Sec.~\ref{ReNA}, we present the numerical analysis
in representative scenarios.
Finally, our conclusions are summarized in
Sec.~\ref{concl}.

\section{Scalar leptoquark setup}\label{model}

We consider a low-energy effective Lagrangian invariant under
the SM gauge group,
$\mathcal{G}_{\mathrm{SM}}
=\textup{SU}(3)_C\times\textup{SU}(2)_L\times\textup{U}(1)_Y$.
Interactions between leptoquarks and fermions are classified
according to the fermion number $F=3B+L$, which takes the
values $0$ or $\pm2$.
Each scalar leptoquark multiplet is characterized by a
definite value of $F$, and its transformation properties
under $\mathcal{G}_{\mathrm{SM}}$ are summarized in
Table~\ref{TBrep}.

\begin{table}[!t]
\caption{Representations of scalar LQ fields under
$\mathcal{G}_{\mathrm{SM}}$. Taken from
Ref.~\cite{PhysRevD.102.075037}.
The electric charges of the scalar leptoquark components
are determined by $Q_{\mathrm{LQ}}=T_3+Y$. The subscript of a leptoquark denotes the dimension of its $\textup{SU}(2)_L$ representation, while a tilde distinguishes multiplets that share the same $\textup{SU}(2)_L$ representation but differ in their
$\textup{U}(1)_Y$ hypercharge.
The column labeled MC indicates whether a given leptoquark admits couplings to both left- and
right-handed charged leptons.}
\label{TBrep}
\begin{ruledtabular}
\begin{tabular}{ccccc}
\textbf{Scalar LQs}
& $\mathcal{G}_{\mathrm{SM}}$
& MC
& $F$
& $Q_{\mathrm{LQ}}$ \\ \hline
$S_3$
& $(\overline{\mathbf{3}},\mathbf{3},1/3)$
& \xmark & $-2$
& $(4/3,1/3,-2/3)$ \\
$R_2$
& $(\mathbf{3},\mathbf{2},7/6)$
& $\checkmark$ & $0$
& $(5/3,2/3)$ \\
$\tilde{R}_2$
& $(\mathbf{3},\mathbf{2},1/6)$
& \xmark & $0$
& $(2/3,-1/3)$ \\
$\tilde{S}_1$
& $(\overline{\mathbf{3}},\mathbf{1},4/3)$
& \xmark & $-2$
& $4/3$ \\
$S_1$
& $(\overline{\mathbf{3}},\mathbf{1},1/3)$
& $\checkmark$ & $-2$
& $1/3$ \\
\end{tabular}
\end{ruledtabular}
\end{table}

We adopt the notation and classification of scalar
leptoquark representations introduced in
Ref.~\cite{DORSNER20161}.
The renormalizable fermionic interactions of scalar
leptoquarks that conserve baryon and lepton
number~\cite{BUCHMULLER1987442} can be organized as
\begin{equation}
\mathcal{L}=\mathcal{L}_{F=-2}+\mathcal{L}_{F=0},
\end{equation}
where $\mathcal{L}$ denotes the leptoquark--fermion
interaction Lagrangian. We impose baryon- and lepton-number conservation and do not include diquark interactions of $S_1$. For the two scalar leptoquarks considered in this work, the relevant fermionic interactions in the gauge basis are given by~\cite{Mandal2019}
\eq{
\mathcal{L}\supset
\bar{Q}^{\,c}i\sigma_2Y_{S_1}LS_1
+\bar{u}_R^{\,c}Z_{S_1}\ell_RS_1
+\bar{\ell}_RY_{R_2}R_2^\dagger Q
-\bar{u}_RZ_{R_2}R_2^Ti\sigma_2L
+\text{h.c.},
}
where $Q$ and $L$ are the left-handed quark and lepton
doublets, while $u_R$ and $\ell_R$ denote the right-handed
up-type quark and charged-lepton singlets.
Here $\sigma_2$ is the second Pauli matrix, and the
superscript $c$ denotes charge conjugation,
$\psi^c\equiv C\bar{\psi}^{\,T}$, with $C$ the
charge-conjugation matrix.
The matrices $Y_{S_1}$, $Z_{S_1}$, $Y_{R_2}$, and
$Z_{R_2}$ are generic Yukawa matrices in flavor space.
Flavor, color, and weak-isospin indices are suppressed.

All interactions are expressed in terms of fermion mass
eigenstates. With the convention
$q_{L,R}^{\mathrm{gauge}}
=U_{L,R}^q q_{L,R}^{\mathrm{mass}}$ ($q=u,d$),
the CKM matrix is $V_{\mathrm{CKM}}=U_L^{u\dagger}U_L^d$.
We adopt the up-type mass-diagonal
basis~\cite{PhysRevD.102.075037,PhysRevD.102.075007},
\eq{
U_L^u=U_R^u=\mathbf{I},
\qquad U_L^d=V_{\mathrm{CKM}}.
}

The up-type-quark vertices below contain no explicit CKM factors, whereas the $\textup{SU}(2)_L$-related vertices involving left-handed down-type quarks do. The charged-lepton loop expressions displayed below contain the up-type-quark contributions from $S_1$ and $R_2^{5/3}$. The down-type-quark interactions are included when evaluating the rare kaon constraints and, for $R_2$, the tree-level coherent $\mu$--$e$ conversion constraint.

With these conventions, the charged-lepton interactions
relevant for the up-type-quark contributions to our
calculation can be written in the compact form
\begin{equation}\label{LLQS1}
\mathcal{L}_{S_1}=\bar{u}_i^{c}\left(\Lambda_L^{ij}P_L+\Lambda_R^{ij}P_R\right)\ell_jS_1+\text{h.c.},
\end{equation}
for the $S_1$ leptoquark, and
\begin{equation}\label{LLQR2}
\mathcal{L}_{R_2^{5/3}}=\bar{u}_i\left(\lambda_L^{ij}P_L+\lambda_R^{ij}P_R\right)\ell_jR_2^{5/3}+\text{h.c.},
\end{equation}
for the component of $R_2$ with electric charge $5/3$.
Here $P_{L,R}=(1\mp\gamma_5)/2$.
The effective couplings $\Lambda_{L,R}^{ij}$ for $S_1$
and $\lambda_{L,R}^{ij}$ for $R_2$ encode the original
Yukawa matrices together with the corresponding flavor
rotations.
The labels $L,R$ refer to the chirality of the
charged-lepton field on which the projectors act.
The indices $i=1,2,3$ label the physical up-type quarks
$u,c,t$, respectively, while $j=1,2,3$ label the charged
leptons $e,\mu,\tau$.\footnote{The Feynman rules for the
interactions between scalar leptoquarks and charged
SM fermions follow from
Eqs.~\eqref{LLQS1} and~\eqref{LLQR2}.}

\section{\texorpdfstring{Low-energy observables and constraints}{Low-energy observables and constraints}}
\label{constraints}

We summarize the low-energy observables used to constrain
the scalar leptoquark benchmarks: the muon anomalous magnetic
moment, radiative and three-body charged-lepton decays, and coherent $\mu$--$e$ conversion. We distinguish the expressions evaluated in the numerical scan
from approximate coupling bounds used only for illustration.
The additional kaon and lepton-flavor-violating $Z$ constraints
are specified in Sec.~\ref{ReNA} and Appendix~\ref{BRZ}.
Together, these observables define the selected constraint set
of our analysis; they do not establish globally allowed
parameter regions \footnote{Constraints from high-energy dilepton tails and single
leptoquark production are not included, nor are loop-level
leptoquark contributions to coherent $\mu$--$e$ conversion}.

For compactness, $\lambda_{L,R}$ denotes the couplings in generic expressions, with the replacement $\lambda_{L,R}\to\Lambda_{L,R}$ understood for $S_1$. The separate up-only, charm-only, and top-only benchmarks retain one row of each up-type Yukawa matrix, as specified in Sec.~\ref{ReNA}.

\subsection{The muon anomalous magnetic moment}\label{DamuPro}
Scalar leptoquarks modify the electromagnetic vertex of charged leptons at one loop. The diagrams in Fig.~\ref{DFPin1} give the following contribution to the anomalous magnetic moment~\cite{Mandal2019}:
\begin{figure}[b!]
    \centering
    \includegraphics[width=0.35\textwidth]{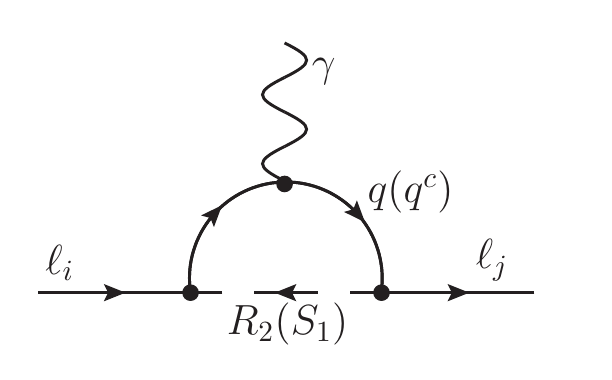}
    \hspace{0.05\textwidth}
    \includegraphics[width=0.35\textwidth]{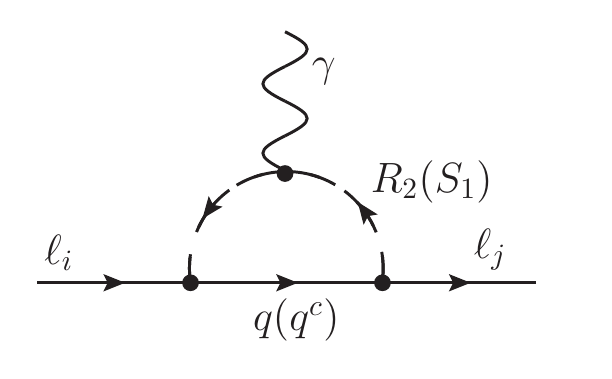}
    \caption{One-loop vertex diagrams contributing to
    $\Delta a_\ell$ through scalar leptoquark exchange.
    For the anomalous magnetic moment, the external leptons
    have the same flavor.}
    \label{DFPin1}
\end{figure}

\begin{equation}\label{Damu}
\Delta a_\ell
=
-\frac{N_c m_\ell^2}{16\pi^2 M_{\mathrm{LQ}}^2}
\sum_q
\bigg[
\left(
|\lambda_L^{q\ell}|^2+|\lambda_R^{q\ell}|^2
\right)\mathcal{A}_1(x_q)
+
\frac{m_q}{m_\ell}
\operatorname{Re}
\left(\lambda_L^{q\ell}\lambda_R^{q\ell*}\right)
\mathcal{A}_2(x_q)
\bigg].
\end{equation}

Here $M_{\mathrm{LQ}}$ is the leptoquark mass,
$N_c=3$ is the number of colors, and the sum runs over
the internal up-type quark flavors associated with
Eqs.~\eqref{LLQS1} and~\eqref{LLQR2}.
The functions $\mathcal{A}_{1,2}(x_q)$ are
\begin{equation}\label{FunAq}
\begin{split}
\mathcal{A}_1(x_q)
&=Q_qF_1(x_q)+Q_{\mathrm{LQ}}F_2(x_q),\\
\mathcal{A}_2(x_q)
&=Q_qF_3(x_q)+Q_{\mathrm{LQ}}F_4(x_q),
\end{split}
\end{equation}
where $x_q=m_q^2/M_{\mathrm{LQ}}^2$.
The electric charges $Q_q$ and $Q_{\mathrm{LQ}}$ are given
in units of the positive elementary charge.
For $S_1$, the internal fermion is a charge-conjugated
up-type quark, so $(Q_q,Q_{\mathrm{LQ}})=(-2/3,1/3)$;
for $R_2^{5/3}$, $(Q_q,Q_{\mathrm{LQ}})=(2/3,5/3)$.
The loop functions are
\begin{equation}
\begin{split}
F_1(x_q)
&=
\frac{
2+3x_q-6x_q^2+x_q^3+6x_q\ln x_q
}{6(1-x_q)^4},
\\
F_2(x_q)
&=
\frac{
1-6x_q+3x_q^2+2x_q^3-6x_q^2\ln x_q
}{6(1-x_q)^4},
\\
F_3(x_q)
&=
\frac{-3+4x_q-x_q^2-2\ln x_q}{(1-x_q)^3},
\\
F_4(x_q)
&=
\frac{1-x_q^2+2x_q\ln x_q}{(1-x_q)^3}.
\end{split}
\end{equation}

Equation~\eqref{Damu} retains the leading nonzero contributions in the external-lepton mass. Additional dependence on $m_\ell^2/M_{\mathrm{LQ}}^2$ in the loop denominators is neglected, while the explicit factors $m_\ell^2/M_{\mathrm{LQ}}^2$ and $m_\ell m_q/M_{\mathrm{LQ}}^2$ are retained.

We adopt $\Delta a_\mu=a_\mu^{\mathrm{exp}}-a_\mu^{\mathrm{SM}}=(38\pm63)\times10^{-11}$~\cite{ALIBERTI20251}. Since this input is compatible with zero within uncertainties, we use it as a consistency constraint without requiring a nonzero leptoquark contribution. At each scan point, Eq.~\eqref{Damu} is evaluated with both same- and mixed-chirality terms, and the result is required to lie within the adopted $1\sigma$ interval. This prescription applies to all three quark benchmarks.

For the illustrative estimates in Tables~\ref{DamRes} and~\ref{B2De}, we adopt the up-type quark masses $m_t=172.57~\mathrm{GeV}$, $m_c=1.273~\mathrm{GeV}$, and $m_u=2.16\times10^{-3}~\mathrm{GeV}$, and the charged-lepton masses $m_\tau=1.77693~\mathrm{GeV}$, $m_\mu=105.66\times10^{-3}~\mathrm{GeV}$, and $m_e=0.511\times10^{-3}~\mathrm{GeV}$, following Ref.~\cite{PhysRevD.110.030001}. We fix $M_{\mathrm{LQ}}=1.7\,\mathrm{TeV}$ as an illustrative benchmark.
Searches by ATLAS and CMS at $\sqrt{s}=13\,\mathrm{TeV}$ constrain leptoquark masses and couplings~\cite{Aad2023,Aad2021LQ,Aad2021LQTau,PhysRevLett.132.061801}, but their bounds depend on the production mechanism,
branching fractions, and flavor structure. The chosen mass is therefore an illustrative scenario.

The term proportional to $m_q\operatorname{Re}(\lambda_L^{q\ell}\lambda_R^{q\ell*})$ can be chirality enhanced in $S_1$ and $R_2^{5/3}$~\cite{Bigaran2019,PhysRevD.102.075037}. Its dominance depends on the relative sizes of the couplings
and is not assumed throughout the scan. When it dominates, Eq.~\eqref{Damu} reduces to
\begin{equation}\label{Daaprox}
\Delta a_\ell
\simeq
-\frac{N_c m_\ell}{16\pi^2M_{\mathrm{LQ}}^2}
\sum_q m_q\mathcal{A}_2(x_q)
\operatorname{Re}
\left(\lambda_L^{q\ell}\lambda_R^{q\ell*}\right).
\end{equation}

Restricting this illustrative approximation to the charm and top contributions gives
\begin{equation}
\begin{split}
&m_c\mathcal{A}_2(x_c)
\operatorname{Re}
\left(\lambda_L^{22}\lambda_R^{22*}\right)
+m_t\mathcal{A}_2(x_t)\operatorname{Re}
\left(\lambda_L^{32}\lambda_R^{32*}\right)\simeq
-\frac{16\pi^2M_{\mathrm{LQ}}^2}{N_cm_\mu}
\Delta a_\mu.
\end{split}
\end{equation}

Only one of these two coupling products is nonzero in each of the charm-only and top-only benchmarks. Evaluating the loop functions at the adopted leptoquark mass yields
\begin{equation}\label{DAres}
\begin{split}
R_2:\quad&
\operatorname{Re}
\left(\lambda_L^{32}\lambda_R^{32*}\right)+0.02392\operatorname{Re}
\left(\lambda_L^{22}\lambda_R^{22*}\right)=(-0.545\pm0.904)\times10^{-3},\\[0.2cm]
S_1:\quad&\operatorname{Re}\left(\Lambda_L^{32}\Lambda_R^{32*}\right)+0.03151\operatorname{Re}\left(\Lambda_L^{22}\Lambda_R^{22*}\right)=(0.803\pm1.332)\times10^{-3}.
\end{split}
\end{equation}

Setting either the charm or top couplings to zero gives the illustrative intervals in Table~\ref{DamRes}. These intervals allow either sign of the coupling products, whereas the numerical scan uses real, positive couplings. They are not imposed as additional cuts and are not extrapolated to the up-only benchmark. For other leptoquark masses, the loop functions must be reevaluated: a quadratic scaling of these coupling bounds is only approximate.
\begin{table}[h!]
\caption{Illustrative $1\sigma$ intervals for the mixed-chirality coupling products at $M_{\mathrm{LQ}}=1.7~\mathrm{TeV}$, obtained from Eq.~\eqref{Daaprox} with one nonzero up-type quark generation. The adopted input is $\Delta a_\mu=(38\pm63)\times10^{-11}$. These intervals are not substitutes for the point-by-point evaluation of Eq.~\eqref{Damu}.}
\label{DamRes}
\begin{ruledtabular}
\begin{tabular}{ccc}
$q$&$R_2$&$S_1$\\\hline
$t$&$\operatorname{Re}(\lambda_L^{32}\lambda_R^{32*})
\in[-14.5,3.59]\times10^{-4}$& $\operatorname{Re}(\Lambda_L^{32}\Lambda_R^{32*})\in[-5.29,21.4]\times10^{-4}$\\
$c$&$\operatorname{Re}(\lambda_L^{22}\lambda_R^{22*})
\in[-6.06,1.50]\times10^{-2}$&$\operatorname{Re}(\Lambda_L^{22}\Lambda_R^{22*})\in[-1.68,6.78]\times10^{-2}$\\
\end{tabular}
\end{ruledtabular}
\end{table}

\subsection{\texorpdfstring{Radiative charged-lepton decays}{Radiative charged-lepton decays}}
\label{Dto2}

\begin{figure}[t!]
    \centering
    \includegraphics[width=0.35\textwidth]{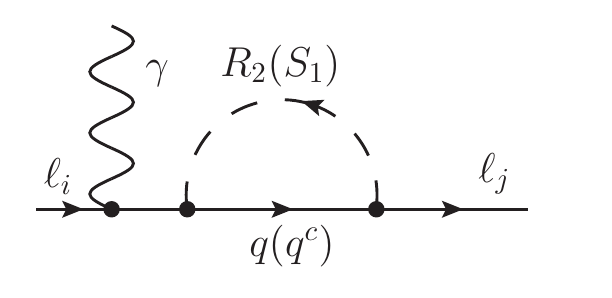}
    \hspace{0.05\textwidth}
    \includegraphics[width=0.35\textwidth]{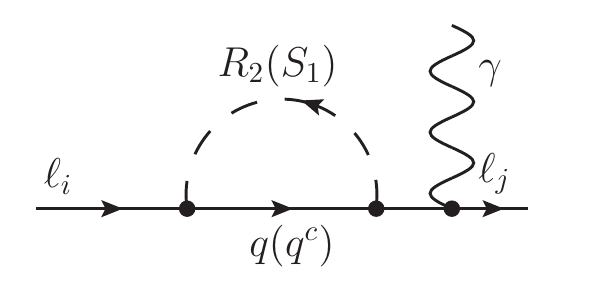}
    \caption{One-loop diagrams with photon emission
    from the external lepton lines contributing to
    $\ell_i\to\ell_j\gamma$.}
    \label{DFPin2}
\end{figure}

The decays $\ell_i\to\ell_j\gamma$ constrain products
of Yukawa couplings involving different charged-lepton flavors.
Their experimental upper limits and projected sensitivities
are collected in Table~\ref{TBlim2d}.
The one-loop contributions are obtained from the vertex
topologies in Fig.~\ref{DFPin1}, with different external
lepton flavors, and the external-line diagrams in
Fig.~\ref{DFPin2}.

\begin{table}[b!]
\caption{Experimental upper limits at $90\%$ C.L. used in the scan and projected sensitivities for radiative CLFV decays. The tau projections are the $50~\mathrm{ab}^{-1}$ estimates in Table~14 of Ref.~\cite{10.1093/ptep/ptz106}.}
\label{TBlim2d}
\begin{ruledtabular}
\begin{tabular}{ccc}
Process&Upper limit ($90\%$ C.L.)&Projected sensitivity\\\hline
$\mathrm{BR}(\mu\to e\gamma)$& $1.5\times10^{-13}$~\cite{Afanaciev2025}&
$6\times10^{-14}$~\cite{Afanaciev2025}\\
$\mathrm{BR}(\tau\to e\gamma)$ &
$3.3\times10^{-8}$~\cite{PhysRevLett.104.021802}&
$1.2\times10^{-8}$~\cite{10.1093/ptep/ptz106}\\
$\mathrm{BR}(\tau\to\mu\gamma)$&$4.2\times10^{-8}$ \cite{Belle:2021ysv}&$5\times10^{-9}$~\cite{10.1093/ptep/ptz106}\\
\end{tabular}
\end{ruledtabular}
\end{table}

Writing $m_i=m_{\ell_i}$ and denoting the total width
of $\ell_i$ by $\Gamma_i$, the branching ratio
is~\cite{Lavoura2003,DORSNER20161}
\begin{equation}\label{BR2d}
\mathrm{BR}(\ell_i\to\ell_j\gamma)
=
\frac{\alpha_{\mathrm{em}}}{4\Gamma_i}
\left(\frac{m_i^2-m_j^2}{m_i}\right)^3
\left(|C_R^{ij}|^2+|C_L^{ij}|^2\right),
\end{equation}
where $\alpha_{em}=Q_{\rm em}^2/4\pi$ and $Q_{\rm em}>0$
is the electromagnetic coupling.
The dipole form factors are
\begin{equation}\label{coe2d}
\begin{split}
C_R^{ij}
&=
\frac{N_c}{32\pi^2M_{\mathrm{LQ}}^2}
\sum_q
\bigg[
\left(
m_i\lambda_L^{qi}\lambda_L^{qj*}
+m_j\lambda_R^{qi}\lambda_R^{qj*}
\right)\mathcal{A}_1(x_q)
+m_q\lambda_R^{qi}\lambda_L^{qj*}
\mathcal{A}_2(x_q)
\bigg],
\\
C_L^{ij}
&=C_R^{ij}(L\leftrightarrow R).
\end{split}
\end{equation}

The loop functions are defined in Eq.~\eqref{FunAq}, and the quark-flavor sum is performed at the amplitude level. Equation~\eqref{BR2d} retains the final-state mass in the
phase-space factor, while Eq.~\eqref{coe2d} uses the leading
external-mass expansion of the loop integrals. The scan applies these expressions at each point, retaining same- and mixed-chirality terms and their interference. In particular, the internal-quark mass alone does not ensure that the mixed-chirality term dominates.

To illustrate the resulting coupling bounds, consider a single quark flavor and assume mixed-chirality dominance. Then
\begin{equation}
C_R^{ij}\simeq
\frac{N_cm_q}{32\pi^2M_{\mathrm{LQ}}^2}
\lambda_R^{qi}\lambda_L^{qj*}
\mathcal{A}_2(x_q),
\end{equation}
with $C_L^{ij}$ obtained by exchanging $L$ and $R$.
For $m_j\ll m_i$, the experimental upper limit implies
\begin{equation}\label{Desto2}
|\lambda_L^{qi}\lambda_R^{qj*}|^2+|\lambda_R^{qi}\lambda_L^{qj*}|^2
<\frac{\Gamma_i}{m_i^3}\mathrm{BR}_{\mathrm{exp}}^{\mathrm{max}}
(\ell_i\to\ell_j\gamma)C_q^2,
\end{equation}
where
\begin{equation}
C_q=\frac{64\pi^2M_{\mathrm{LQ}}^2}{N_cm_q\sqrt{\alpha_{\mathrm{em}}}\mathcal{A}_2(x_q)}.
\end{equation}

The coupling combination entering this bound is denoted by
\begin{equation}
|\lambda_{LR}^{qij}|^2\equiv|\lambda_L^{qi}\lambda_R^{qj*}|^2
+|\lambda_R^{qi}\lambda_L^{qj*}|^2.
\end{equation}

For $S_1$, the corresponding combination is $|\Lambda_{LR}^{qij}|^2$.

Using the limits in Table~\ref{TBlim2d}, we obtain the illustrative top- and charm-quark bounds in Table~\ref{B2De} at $M_{\mathrm{LQ}}=1.7\,\mathrm{TeV}$. In addition to the masses specified above, the inputs are $\alpha_{\mathrm{em}}^{-1}=137.036$, $\tau_\mu=2.1969811\times10^{-6}~\mathrm{s}$,
and $\tau_\tau=290.3\times10^{-15}~\mathrm{s}$, with $\Gamma_i=\hbar/\tau_i$ and $\hbar=6.582119569\times10^{-25}\,\mathrm{GeV\,s}$. The lepton inputs follow Ref.~\cite{PhysRevD.110.030001}.

\begin{table*}[t!]
\caption{Bounds on mixed-chirality
coupling combinations at
$M_{\mathrm{LQ}}=1.7\,\mathrm{TeV}$, derived from
Eq.~\eqref{Desto2} under mixed-chirality dominance.
For related bounds in other scalar leptoquark models,
see Ref.~\cite{Mandal2019}.}
\label{B2De}
\begin{ruledtabular}
\begin{tabular}{cccc}
Leptoquark &
$\mu\to e\gamma$ &
$\tau\to e\gamma$ &
$\tau\to\mu\gamma$
\\ \hline
$R_2$ &
$|\lambda_{LR}^{321}|^2<1.92\times10^{-15}$ &
$|\lambda_{LR}^{331}|^2<6.7\times10^{-7}$ &
$|\lambda_{LR}^{332}|^2<8.6\times10^{-7}$\\
&$|\lambda_{LR}^{221}|^2<3.35\times10^{-12}$ &
$|\lambda_{LR}^{231}|^2<1.2\times10^{-3}$ &
$|\lambda_{LR}^{232}|^2<1.5\times10^{-3}$\\\hline$S_1$&
$|\Lambda_{LR}^{321}|^2<4.17\times10^{-15}$&
$|\Lambda_{LR}^{331}|^2<1.5\times10^{-6}$&
$|\Lambda_{LR}^{332}|^2<1.9\times10^{-6}$\\&
$|\Lambda_{LR}^{221}|^2<4.20\times10^{-12}$&
$|\Lambda_{LR}^{231}|^2<1.5\times10^{-3}$&
$|\Lambda_{LR}^{232}|^2<1.9\times10^{-3}$\\
\end{tabular}
\end{ruledtabular}
\end{table*}

The strongest bounds in Table~\ref{B2De} arise from
$\mu\to e\gamma$, particularly for the top-quark contribution. For this decay, the upper bounds on the squared coupling
combinations are stronger for $R_2$ than for $S_1$ by factors
of approximately $2.17$ and $1.25$ for the top and charm
contributions, respectively. At fixed experimental inputs, Eq.~\eqref{Desto2} scales as $M_{\mathrm{LQ}}^4/|\mathcal{A}_2(x_q)|^2$; a pure quartic scaling neglects the mass dependence of the loop functions. These approximate bounds illustrate the coupling dependence and do not replace Eqs.~\eqref{BR2d} and~\eqref{coe2d} in any of the three benchmarks analyzed in Sec.~\ref{ReNA}.

\subsection{\texorpdfstring{Three-body charged-lepton decays}{Three-body charged-lepton decays}}\label{threebodyconstraints}
The decays $\ell_i^-\to\ell_j^-\ell_j^-\ell_j^+$
provide complementary constraints on the Yukawa couplings.
For $\mu^-\to e^-e^-e^+$, SINDRUM reports
$\mathrm{BR}(\mu\to3e)<1.0\times10^{-12}$~\cite{BELLGARDT19881},
which constrains the electron--muon coupling combinations.
The decays $\tau^-\to e^-e^-e^+$ and
$\tau^-\to\mu^-\mu^-\mu^+$ constrain combinations
involving the tau flavour. These channels receive photon-penguin, $Z$-penguin, and box contributions. We retain the photon vector and dipole form factors, the vector $Z$ form factors, and the box amplitudes, including their interference. The $Z$ dipole form factors are neglected in the adopted approximation~\cite{Crivellin2021,delAguila2009,Arnan2019}. Final-state masses are neglected except in the logarithm regulating the photon contribution. The branching ratio and form factors are given in Eqs.~\eqref{li3lj}--\eqref{FacFB3l} of Appendix~\ref{Oproc}.

The remaining trilepton tau channels included in the constraint set are discussed in Sec.~\ref{RDecay} and Appendix~\ref{Oproc}. Their experimental limits are collected in Table~\ref{Decay3B}, and the numerical implementation is described in Sec.~\ref{ReNA}.

\subsection{\texorpdfstring{Coherent $\mu$--$e$ conversion}{Coherent mu-e conversion}}\label{mueconversion}

Scalar leptoquarks can mediate coherent $\mu$--$e$ conversion in nuclei at tree level. We retain the light-quark scalar and vector contributions. The conversion rates are calculated using the nuclear matching procedure described in~\cite{PhysRevD.66.096002} and are normalized to the muon capture rate $\omega_{\mathrm{capt}}$. The nuclear overlap integrals $S^{(p,n)}$ and $V^{(p,n)}$ are written in the convention in which $m_\mu^5$ is extracted explicitly.

For the full $R_2$ doublet with degenerate component masses,
the conversion ratio is 
\begin{equation}\label{CRR2}
\mathcal{R}_{R_2}\left(\mu-e,(A,Z)\right)
=\frac{m_{\mu}^5}{4M_{\mathrm{LQ}}^4\omega_{\mathrm{capt}}}
\left(|\mathcal{F}_L^{R_2}|^2+|\mathcal{F}_R^{R_2}|^2\right),
\end{equation}
where the factors containing the Yukawa couplings
and nuclear matrix elements are
\begin{equation}\label{FFCRR2}
\begin{split}
\mathcal{F}_L^{R_2}
&=\left[S^{(p)}G_S^{(u,p)}+S^{(n)}G_S^{(u,n)}\right]
\lambda_L^{u1*}\lambda_R^{u2}
+\left(2V^{(p)}+V^{(n)}\right)\lambda_L^{u1*}\lambda_L^{u2},\\
\mathcal{F}_R^{R_2}
&=\mathcal{F}_L^{R_2}(L\leftrightarrow R)
+\left(V^{(p)}+2V^{(n)}\right)\eta_R^{d1*}\eta_R^{d2},\\
\eta_R^{d\ell}&=\sum_{q=u,c,t}V_{qd}^{*}\lambda_R^{q\ell}.
\end{split}
\end{equation}

Here $1$ and $2$ denote the electron and muon, respectively, $d$ is the first-generation down quark, and $V_{qd}$ is an element of $V_{\mathrm{CKM}}$. The up-quark terms arise from $R_2^{5/3}$ exchange, whereas the down-quark term arises from $R_2^{2/3}$ exchange. For $S_1$, the corresponding expressions are
\begin{equation}\label{CRS1}
\begin{split}
\mathcal{R}_{S_1}(\mu-e,(A,Z))
&=\frac{m_\mu^5}{4M_{\mathrm{LQ}}^4\omega_{\mathrm{capt}}}
\left(|\mathcal{F}_L^{S_1}|^2+|\mathcal{F}_R^{S_1}|^2\right),\\
\mathcal{F}_L^{S_1}
&=\left[S^{(p)}G_S^{(u,p)}+S^{(n)}G_S^{(u,n)}\right]
\Lambda_L^{u1*}\Lambda_R^{u2}
-\left(2V^{(p)}+V^{(n)}\right)\Lambda_L^{u1*}\Lambda_L^{u2},\\
\mathcal{F}_R^{S_1}&=\mathcal{F}_L^{S_1}(L\leftrightarrow R).
\end{split}
\end{equation}

The moduli squared retain interference among the
displayed scalar and vector contributions.

We impose the SINDRUM~II upper limit for gold, $\mathcal{R}(\mu-e,\mathrm{Au})<7\times10^{-13}$, at the $90\%$ confidence level~\cite{Bertl2006}. For the numerical analysis, we use $\omega_{\mathrm{capt}}=8.603\times10^{-18}~\mathrm{GeV}$, corresponding to the gold capture rate $13.07\times10^6~\mathrm{s}^{-1}$ tabulated in Ref.~\cite{PhysRevD.66.096002}. The nuclear overlap integrals are taken to be $S^{(p)}=0.0523$, $S^{(n)}=0.0781$, $V^{(p)}=0.0866$, and $V^{(n)}=0.129$, following Ref.~\cite{HEECK2022115833}. For the scalar nucleon form factors, we adopt $G_S^{(u,p)}=5.1$ and $G_S^{(u,n)}=4.3$ from Ref.~\cite{PhysRevD.66.096002}.

In the up-only benchmark, up-quark contributions are present in both representations, together with the CKM-related down-quark contribution for $R_2$. In the charm-only and top-only benchmarks, the up-quark couplings vanish, so the displayed tree-level contribution vanishes for $S_1$. For $R_2$, the single-generation relation $\eta_R^{d\ell}=V_{qd}^{*}\lambda_R^{q\ell}$ implies
$\eta_R^{d1*}\eta_R^{d2}=|V_{qd}|^2\lambda_R^{q1*}\lambda_R^{q2}$, with $q=c,t$; the remaining conversion ratio therefore carries a factor $|V_{qd}|^4$.

\section{\texorpdfstring
{Box-induced $\tau^-\to\ell_i^-\ell_i^-\ell_j^+$ ($i\neq j$) decays}
{Box-induced tau- -> li- li- lj+ (i != j) decays}
}\label{RDecay}

\begin{figure}[b!]
    \centering
    \includegraphics[width=0.35\textwidth]{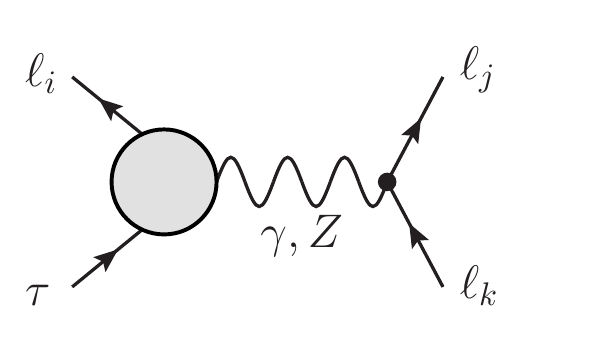}
    \hspace{0.05\textwidth}
    \includegraphics[width=0.3\textwidth]{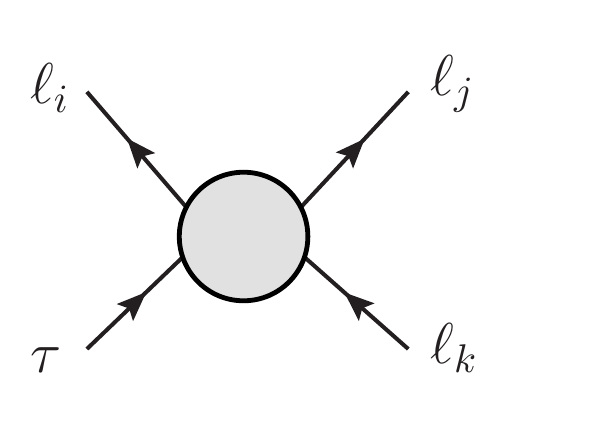}
    \caption{Generic penguin and box diagrams contributing to
    $\tau^-\to\ell_i^-\ell_j^-\ell_k^+$.
    Penguin contributions require a same-flavor
    lepton--antilepton pair in the final state.}
    \label{DFg}
\end{figure}

We now consider three-body CLFV tau decays. There are six
independent channels of the form
$\tau^-\to\ell_i^-\ell_j^-\ell_k^+$ with
$\ell_{i,j,k}=e,\mu$.
These modes can be conveniently classified into two categories
according to their flavor structure.

The first class corresponds to decays of the type $\tau^-\to\ell_i^-\ell_j^-\ell_j^+$, with $\ell_i,\ell_j\in\{e,\mu\}$, including the case $i=j$, namely $\tau^-\to e^-e^-e^+$, $\tau^-\to\mu^-\mu^-\mu^+$, $\tau^-\to e^-\mu^-\mu^+$, and $\tau^-\to\mu^-e^-e^+$. In these channels, the same-flavor lepton--antilepton pair $\ell_j^-\ell_j^+$ can originate from an off-shell photon or $Z$ boson. Consequently, they receive contributions from both penguin diagrams, mediated by $\gamma$ and $Z$ exchange, and box diagrams at the one-loop level. The generic penguin and box diagrams contributing to these processes are shown in Fig.~\ref{DFg}.

\begin{figure}[b!]
    \centering
    \includegraphics[width=0.3\textwidth]{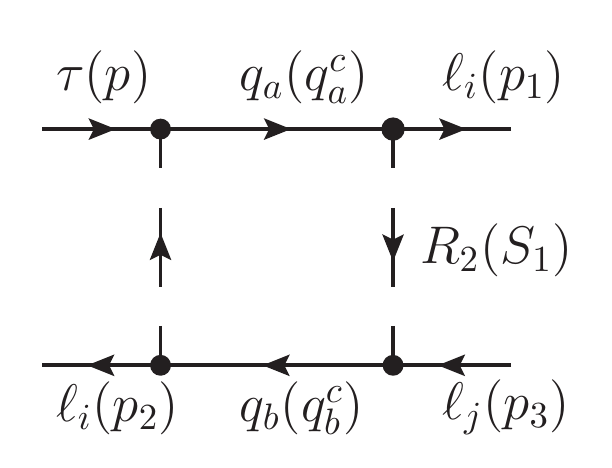}
    \hspace{0.05\textwidth}
    \includegraphics[width=0.3\textwidth]{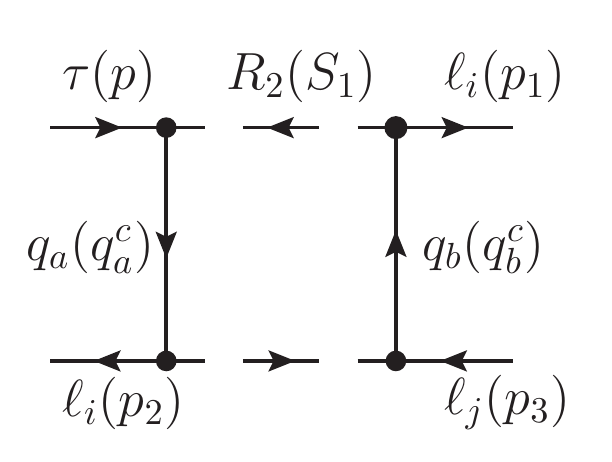}
    \caption{Box diagrams contributing to the three-body decays
    $\tau^-\to\ell_i^-\ell_i^-\ell_j^+$ with $i\neq j$.
    The internal up-type quarks are denoted by $q_a$ and $q_b$,
    with $a,b=1,2,3$ corresponding to $u,c,t$.}
    \label{DBox}
\end{figure}

The second class consists of the double-flavor-changing channels
$\tau^-\to\ell_i^-\ell_i^-\ell_j^+$ with $i\neq j$, namely
$\tau^-\to e^-e^-\mu^+$ and $\tau^-\to\mu^-\mu^-e^+$. In these decays, the final state does not contain a same-flavor lepton--antilepton pair, and therefore contributions from photon- or $Z$-mediated penguin diagrams are absent at one loop. As a result, these processes are generated exclusively by box diagrams in the models considered here (see Fig.~\ref{DBox}).

While previous studies have mainly focused on the first class of channels, we also investigate the double-flavor-changing modes. Their study probes complementary combinations of leptoquark Yukawa couplings and characterizes the box contributions without interference from penguin amplitudes. Their relevance therefore lies in elucidating the flavor and chiral structure of the interactions and quantifying the suppression of these channels. The experimental upper limits and projected sensitivities for CLFV three-body tau decays are summarized in
Table~\ref{Decay3B}.
\begin{table}[t!]
\caption{Experimental upper limits at 90\% C.L. and projected
sensitivities for CLFV three-body tau decays. The projections assume $50~\mathrm{ab}^{-1}$.}
\label{Decay3B}
\begin{ruledtabular}
\begin{tabular}{ccc}
Process & Upper limit (90\% C.L.) & Projected sensitivity \\
\hline
$\text{BR}(\tau^-\to\mu^-\mu^-\mu^+)$
&$1.9\times10^{-8}$~\cite{Adachi2024}
& $3.6\times10^{-10}$~\cite{Banerjee2022} \\
$\text{BR}(\tau^-\to e^-e^-e^+)$
& $2.5\times10^{-8}$~\cite{Adachi2025}
& $4.7\times10^{-10}$~\cite{Banerjee2022} \\
$\text{BR}(\tau^-\to e^-\mu^-\mu^+)$
& $2.4\times10^{-8}$~\cite{Adachi2025}
& $4.5\times10^{-10}$~\cite{Banerjee2022} \\
$\text{BR}(\tau^-\to\mu^-e^-e^+)$
& $1.6\times10^{-8}$~\cite{Adachi2025}
& $2.9\times10^{-10}$~\cite{Banerjee2022} \\
$\text{BR}(\tau^-\to e^-e^-\mu^+)$
&$1.5\times10^{-8}$~\cite{HAYASAKA2010139}
& $2.3\times10^{-10}$~\cite{Banerjee2022} \\
$\text{BR}(\tau^-\to\mu^-\mu^-e^+)$
& $1.3\times10^{-8}$~\cite{Adachi2025}
& $2.6\times10^{-10}$~\cite{Banerjee2022} \\
\end{tabular}
\end{ruledtabular}
\end{table}

The loop integrals entering the box amplitudes were evaluated
using \texttt{Package-X}~\cite{PATEL2015276}. The loop integrals are evaluated in the limit of vanishing external momenta.
In this approximation, the amplitudes reduce to local
four-fermion interactions in the low-energy effective
theory~\cite{delAguila2009}. The external momenta are retained in the spinor structures and in the three-body phase-space integration. The final-state charged-lepton masses are neglected
in the branching-ratio expression below.

We assign momenta according to $\tau^-(p)\to\ell_i^-(p_1)\ell_i^-(p_2)\ell_j^+(p_3)$. The contributions related by the exchange of the two identical final-state leptons are combined with the relative minus sign required by Fermi statistics. Fierz rearrangements, see Appendix~\ref{Amp}, then allow the resulting amplitude to be expressed using a common ordering of the external
spinors:
\begin{equation}
\begin{split}
\mathcal{M}_{\text{Box}}&=\sum_{X,Y=L,R}\left(S_{XY}[\bar{u}_1P_Xu_p][\bar{u}_2P_Yv_3]+V_{XY}[\bar{u}_1\gamma^\alpha P_Xu_p]
[\bar{u}_2\gamma_\alpha P_Yv_3]\right)\\[0.2cm]
&+\sum_{X=L,R}T_{XX}
[\bar{u}_1\sigma^{\alpha\beta}P_Xu_p]
[\bar{u}_2\sigma_{\alpha\beta}P_X v_3],
\end{split}
\end{equation}
where $\sigma^{\alpha\beta}=i[\gamma^\alpha,\gamma^\beta]/2$.
The scalar, vector, and tensor form factors have mass
dimension $-2$ and refer to the amplitude after combining
the identical-lepton exchange contributions.
In this convention, the form factors satisfy
\begin{equation}
T_{LL}=-\frac{1}{4}S_{LL},
\qquad
T_{RR}=-\frac{1}{4}S_{RR},
\qquad
V_{LR}=-\frac{1}{2}S_{LR},
\qquad
V_{RL}=-\frac{1}{2}S_{RL}.
\end{equation}

These relations are used before performing the phase-space integration. Thus, the tensor and mixed-chirality vector contributions are already incorporated in the expression below.

After integrating the differential decay width over the three-body phase space, the branching ratio takes the same functional form for $S_1$ and $R_2$:
\eq{\label{BRF0}
\text{BR}=
\frac{m_{\tau}^5}{512\pi^3\Gamma_{\tau}}
\bigg\{
\frac{1}{6}\left(|S_{LL}|^2+|S_{RR}|^2+|V_{LL}|^2+|V_{RR}|^2\right)
+\frac{1}{12}\left(|S_{LR}|^2+|S_{RL}|^2\right)
\bigg\}.
}

Here, $L,R$ specify the chiral projectors in the corresponding fermion bilinears, and $\Gamma_\tau$ is the total tau decay width. The explicit form factors in terms of leptoquark Yukawa couplings and Passarino--Veltman loop functions are collected in Appendix~\ref{App:BoxDetails}.

For the $S_1$ model, the Yukawa couplings are denoted by $\Lambda_{L,R}^{q\ell}$ instead of $\lambda_{L,R}^{q\ell}$. Equation~\eqref{BRF0} is evaluated using the form factors appropriate to each representation, with the conventions specified in Appendix~\ref{App:BoxDetails}. Its common functional form does not imply identical phenomenological predictions: the constraints imposed on the Yukawa couplings depend on the leptoquark representation.

\section{Numerical analysis}\label{ReNA}
Building upon the analytical expression derived in the previous
section for the $R_2$ and $S_1$ leptoquark representations,
we begin with a simplified study aimed at illustrating
the qualitative behavior of Eq.~\eqref{BRF0}.
To this end, we assume the Yukawa couplings
$\lambda_{L,R}^{q\ell}$ to be real and flavor- and
chirality-universal, $\lambda_{L,R}^{q\ell}=\lambda$,
which considerably simplifies the full expressions.
Here and below, $\lambda_{L,R}^{q\ell}$ denotes the
couplings generically; for $S_1$, it is replaced by
$\Lambda_{L,R}^{q\ell}$. This universal-coupling example is distinct from the up-only, charm-only, and top-only benchmarks introduced below and is not subjected to the full set of phenomenological constraints used in the subsequent scan.

In the regime $M_{\mathrm{LQ}}^2\gg m_q^2$, and for the universal couplings assumed here, the contributions proportional to $m_{q_a}m_{q_b}D_0$ are suppressed relative to those involving $D_{00}$; see Appendix~\ref{App:BoxDetails}. Neglecting these contributions gives

\begin{equation}\label{BRmax}
\text{BR}_{\lambda}=\frac{m_\tau^5}{512\pi^3\Gamma_\tau}
\left|\frac{N_c\lambda^4}{8\sqrt{2}\pi^2}\sum_{a,b}D_{00}(x_a,y_b)\right|^2.
\end{equation}

The sums in this illustrative universal-coupling example run over the three up-type quark generations. The approximation concerns the relative importance of the loop structures and does not establish equivalent phenomenology for the two representations.
\begin{figure}[!t]
\centering
\includegraphics[scale=1]{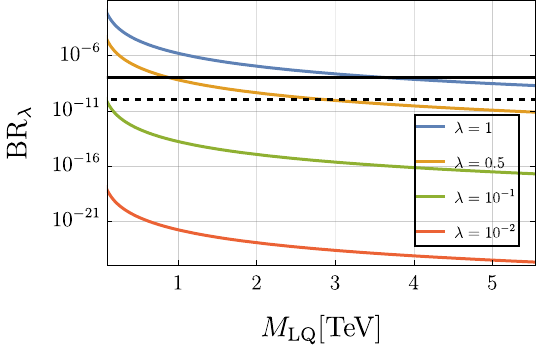}
\caption{Branching ratio for
$\tau^-\to\ell_i^-\ell_i^-\ell_j^+$ with real,
universal Yukawa couplings.
The solid and dashed black lines indicate the
approximate scales of the current upper limits and
projected sensitivities, respectively. The blue, orange, green, and red curves correspond to $\lambda=1,0.5,0.1,0.01$, respectively. The phenomenological constraints imposed in the
subsequent numerical scan are not applied in this
illustrative example.}\label{BRlim}
\end{figure}

We explore Eq.~\eqref{BRmax} by varying $\lambda$ within $(0,1]$ and $M_{\mathrm{LQ}}$ in the TeV range. The results are displayed in Fig.~\ref{BRlim}. The strong dependence on the universal coupling, $\text{BR}_{\lambda}\propto\lambda^8$, illustrates how
sizable branching ratios can arise before imposing the other phenomenological constraints. This example does not establish that experimentally accessible rates survive those constraints.

We now relax the assumption of universal Yukawa couplings. For each scalar leptoquark representation, the charged-lepton couplings to up-type quarks comprise a $3\times3$ matrix per chirality, $\lambda_{L,R}^{q\ell}$, with $q,\ell=1,2,3$. Restricting these couplings to real values leaves 18 independent parameters before imposing additional assumptions about their flavor structure.

CLFV processes constrain products of Yukawa couplings involving
different charged-lepton flavors~\cite{PhysRevD.108.095027}. In each benchmark, we allow nonzero entries $\lambda_{L,R}^{a\ell}$ in a single row $a$ of the up-type Yukawa matrices, with $\ell=1,2,3$ spanning all three charged-lepton flavors. The choices $a=1,2,3$ define the up-only, charm-only, and top-only benchmarks, respectively.

The choice of row is an assumption about the Yukawa flavor
structure. In particular, the up-quark box terms proportional
to $D_{00}$ can survive in the limit $m_u\to 0$, since at
vanishing external momenta and fixed $M_{\mathrm{LQ}}$,
\begin{equation}
D_{00}(x_q,x_q)
\xrightarrow{x_q\to0}
-\frac{1}{4M_{\mathrm{LQ}}^2},
\qquad
x_q=\frac{m_q^2}{M_{\mathrm{LQ}}^2}.
\end{equation}

Therefore, each scenario contains six independent real Yukawa couplings at fixed leptoquark mass. For a fixed up-type quark $q$, and retaining only the $D_{00}$ contributions,
Eq.~\eqref{BRF0} reduces to
\begin{equation}\label{BReq}
\begin{split}
\text{BR}\left(\tau^-\to\ell_i^-\ell_i^-\ell^+_j\right)=\frac{m_\tau^5}{512\pi^3\Gamma_\tau}
\left(\frac{N_cD_{00}^{q}}{16\pi^2\sqrt{3}}\right)^2&\bigg\{2\left[\lambda_L^{q3}\lambda_L^{qj}(\lambda_L^{qi})^2\right]^2+2\left[\lambda_R^{q3}\lambda_R^{qj}(\lambda_R^{qi})^2\right]^2\\
&+\left[\lambda_L^{qi}\lambda_L^{qj}\lambda_R^{q3}\lambda_R^{qi}\right]^2+\left[\lambda_L^{q3}\lambda_L^{qi}\lambda_R^{qi}\lambda_R^{qj}\right]^2\bigg\}.
\end{split}
\end{equation}

Here $D_{00}^{q}\equiv D_{00}(x_q)$ and the lepton index $3$ denotes the tau. The following factorization assumes strictly positive Yukawa couplings and applies within this
$D_{00}$ approximation.

Introducing the auxiliary ratios
$r_{q\ell}\equiv\lambda_L^{q\ell}/\lambda_R^{q\ell}$,
we write
\begin{equation}
\text{BR}\left(\tau^-\to\ell_i^-\ell_i^-\ell_j^+\right)=
\frac{m_\tau^5}{512\pi^3\Gamma_\tau}r_q\lambda_{RL}^{q3ji^2},
\end{equation}
where
\begin{equation}
\lambda_{RL}^{q3ji^2}=\lambda_R^{q3}\lambda_L^{q3}\lambda_R^{qj}\lambda_L^{qj}\left(\lambda_R^{qi}\lambda_L^{qi}\right)^2
\end{equation}
and
\begin{equation}
r_q=\left(\frac{N_cD_{00}^{q}}{16\pi^2\sqrt{3}}\right)^2
\frac{2+r_{q3}^2r_{qi}^2+r_{qj}^2r_{qi}^2
+2r_{q3}^2r_{qj}^2r_{qi}^4}{r_{q3}r_{qj}r_{qi}^2}.
\end{equation}

Unlike the dimensionless ratios $r_{q\ell}$, the function $r_q$ has mass dimension $-4$. For positive chiral ratios, its inverse satisfies
\begin{equation}
\begin{aligned}
r_q^{-1}&=\left(\frac{16\pi^2\sqrt{3}}{N_cD_{00}^{q}}\right)^2
\left[2\left(r_{q3}r_{qj}r_{qi}^2+\frac{1}{r_{q3}r_{qj}r_{qi}^2}\right)
+\frac{r_{q3}}{r_{qj}}+\frac{r_{qj}}{r_{q3}}\right]^{-1}
\leq\frac{1}{6}\left(\frac{16\pi^2\sqrt{3}}{N_cD_{00}^{q}}\right)^2.
\end{aligned}
\end{equation}
This follows from $x+x^{-1}\geq2$ for $x>0$. Equality holds when
\begin{equation}
r_{q3}=r_{qj},\qquad r_{qi}=\frac{1}{r_{q3}}.
\end{equation}
The choice $r_{q3}=r_{qj}=r_{qi}=1$ is one member of this continuous family of maximizing configurations. Using this maximum and an experimental upper limit $\mathrm{BR}_{\mathrm{exp}}$, we obtain
\begin{equation}\label{desLam}
\lambda_{RL}^{q3ji^2}<\frac{512\pi^3\Gamma_\tau}{m_\tau^5}
\left(\frac{8\sqrt{2}\pi^2}{N_cD_{00}^{q}}\right)^2
\mathrm{BR}_{\mathrm{exp}}\left(\tau^-\to\ell_i^-\ell_i^-\ell_j^+\right).
\end{equation}

The function $r_q$ contains the dependence on the chiral
ratios and the loop function, while $\lambda_{RL}^{q3ji^2}$ collects the product of eight Yukawa couplings. For fixed chiral ratios, the experimental branching-ratio limit gives a corresponding upper bound on this product. Equation~\eqref{desLam} gives the weakest such bound over all positive chiral ratios. It is a necessary condition derived from the target decay alone, within the stated approximation; it does not guarantee that a point satisfies the remaining phenomenological constraints.
\begin{figure}[!t]
\centering
\begin{tabular}{cc}
\includegraphics[width=0.455\linewidth]{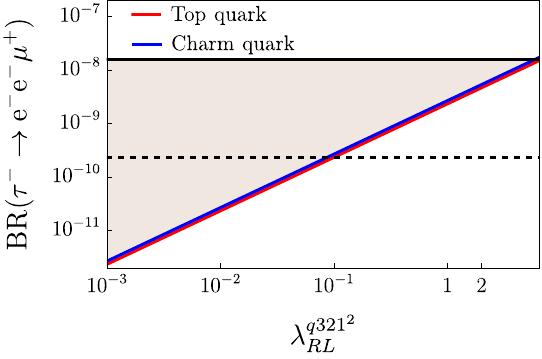}&
\hspace{0.5cm}\includegraphics[width=0.455\linewidth]{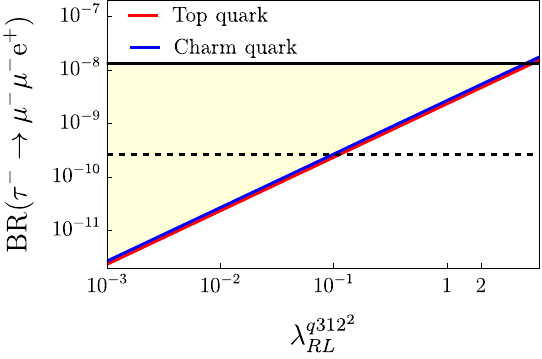}
\end{tabular}
\caption{Minimum branching ratio at fixed $\lambda_{RL}^{q3ji^2}$ in the $D_{00}$ approximation at $M_{\mathrm{LQ}}=1.7~\mathrm{TeV}$, for the top-only (red) and charm-only (blue) benchmarks. The left and right panels show $\tau^-\to e^-e^-\mu^+$ and $\tau^-\to\mu^-\mu^-e^+$, respectively. Solid and dashed black lines denote the Belle~II upper limit and projected sensitivity specified in Table~\ref{Decay3B}. The shading extends from the blue curve to the solid line. The intersection of either colored curve with a horizontal line gives the corresponding coupling-product bound in Eq.~\eqref{desLam}. No other constraint is imposed here.}\label{reg3D}
\end{figure}

For fixed quark masses and $M_{\mathrm{LQ}}\gg m_q$,
$D_{00}^{q}$ scales approximately as $M_{\mathrm{LQ}}^{-2}$. The upper bound in Eq.~\eqref{desLam} therefore increases approximately as $M_{\mathrm{LQ}}^4$. Perturbative limits on the individual couplings also bound their product, but these considerations alone do not select a preferred leptoquark mass range. Figure~\ref{reg3D} uses the Belle~II limits and
projected sensitivities in Table~\ref{Decay3B},
with $M_{\mathrm{LQ}}=1.7~\mathrm{TeV}$ as discussed
in Sec.~\ref{DamuPro}. For this mass, the charm-only benchmark yields a stronger bound on the coupling product than the top-only
benchmark, owing to the quark-mass dependence of
$D_{00}^{q}$. For a fixed quark benchmark, the plotted limit
on $\tau^-\to\mu^-\mu^-e^+$ gives the stronger bound,
since $1.3\times10^{-8}<1.6\times10^{-8}$. For the projected sensitivities, the ordering is reversed: the $e^-e^-\mu^+$ channel gives the stronger prospective bound.

We now turn to the numerical analysis of Eq.~\eqref{BRF0}, evaluating its form factors for a single generation at a time. The equivalent expansion is given in Eq.~\eqref{BRF0num} in Appendix~\ref{App:BoxDetails}. The preceding $D_{00}$ approximation serves to illustrate the coupling dependence and is not an additional constraint on the numerical scan.

The analysis examines the dependence of the branching ratios on the Yukawa couplings in both representations. We restrict the nonzero couplings to be real and positive and impose the perturbative upper bound~\cite{Allwicher2021}
\begin{equation}\label{LimPer}
0<\lambda_{L,R}^{q\ell}<\sqrt{4\pi}.
\end{equation}

The results are therefore restricted to these positive, real, single-generation benchmarks and do not cover general relative signs, complex phases, or simultaneous contributions from different up-type quark generations. For each fixed quark generation, $q=1,2,3$, the six free couplings are
\begin{equation}
\left\{
\lambda_L^{q1},\lambda_R^{q1},
\lambda_L^{q2},\lambda_R^{q2},
\lambda_L^{q3},\lambda_R^{q3}
\right\}.
\end{equation}

The leptoquark mass is fixed to the illustrative value
$M_{\mathrm{LQ}}=1.7~\mathrm{TeV}$. For each benchmark the six nonzero couplings are sampled independently and uniformly in $\log_{10}\lambda$ between $\log_{10}(10^{-12})$ and $\log_{10}\sqrt{4\pi}$. The scan uses $m_t=172.56~\mathrm{GeV}$ and $\tau_\mu=2.197\times10^{-6}~\mathrm{s}$, compared with $172.57~\mathrm{GeV}$ and $2.1969811\times10^{-6}~\mathrm{s}$ in the illustrative estimates. The electroweak inputs are $M_Z=91.1879~\mathrm{GeV}$, $\Gamma_Z=2.4955~\mathrm{GeV}$ and $s_W^2=0.2229$.

At each scan point, we require the predicted $\Delta a_\mu$ to lie within the adopted $1\sigma$ interval and impose the experimental upper limits on $\ell_i\to\ell_j\gamma$ (Table~\ref{TBlim2d}), $\mu\to3e$, $\tau^-\to\ell_i^-\ell_j^-\ell_k^+$ (Table~\ref{Decay3B}), $Z\to\ell_i^\pm\ell_j^\mp$ (Table~\ref{TBZLFV}), together with the tree-level coherent $\mu$--$e$ conversion constraint in Sec.~\ref{mueconversion}, for each of the three benchmarks. These observables are evaluated using
Eqs.~\eqref{Damu}, \eqref{BR2d}, \eqref{li3lj}, \eqref{lilkljlj}, and \eqref{BRZLCFV}, together with Eq.~\eqref{BRF0} for the exclusively box-induced tau decays, Eqs.~\eqref{CRR2}--\eqref{CRS1} for conversion, and with the form factors appropriate to each representation. In particular, the radiative-decay constraints include both same- and mixed-chirality terms and their interference.

Additional constraints involve the down-type-quark interactions related to the selected Yukawa couplings through CKM mixing. For $S_1$, we include $K^+\to\pi^+\nu\bar{\nu}$ using Refs.~\cite{PhysRevD.99.055007,PhysRevD.108.095027}. We require the new-physics shift in the branching ratio, including interference with the SM, to lie between $-1.27\times10^{-10}$ and $3.13\times10^{-10}$. For $R_2$, we assume degenerate doublet components,
\begin{equation}
M_{R_2^{5/3}}=M_{R_2^{2/3}}=M_{R_2},
\end{equation}
thereby eliminating the mass-splitting contribution to the electroweak oblique parameter $T$~\cite{DORSNER20161}.
We impose the constraints from $K_L^0\to e^+e^-$, $K_L^0\to\mu^+\mu^-$, and $K_L^0\to e^\pm\mu^\mp$, which receive
$R_2^{2/3}$ contributions, following Ref.~\cite{Mandal2019}.

\begin{figure}[!t]
\centering
\begin{tabular}{cc}
\includegraphics[width=0.455\linewidth]{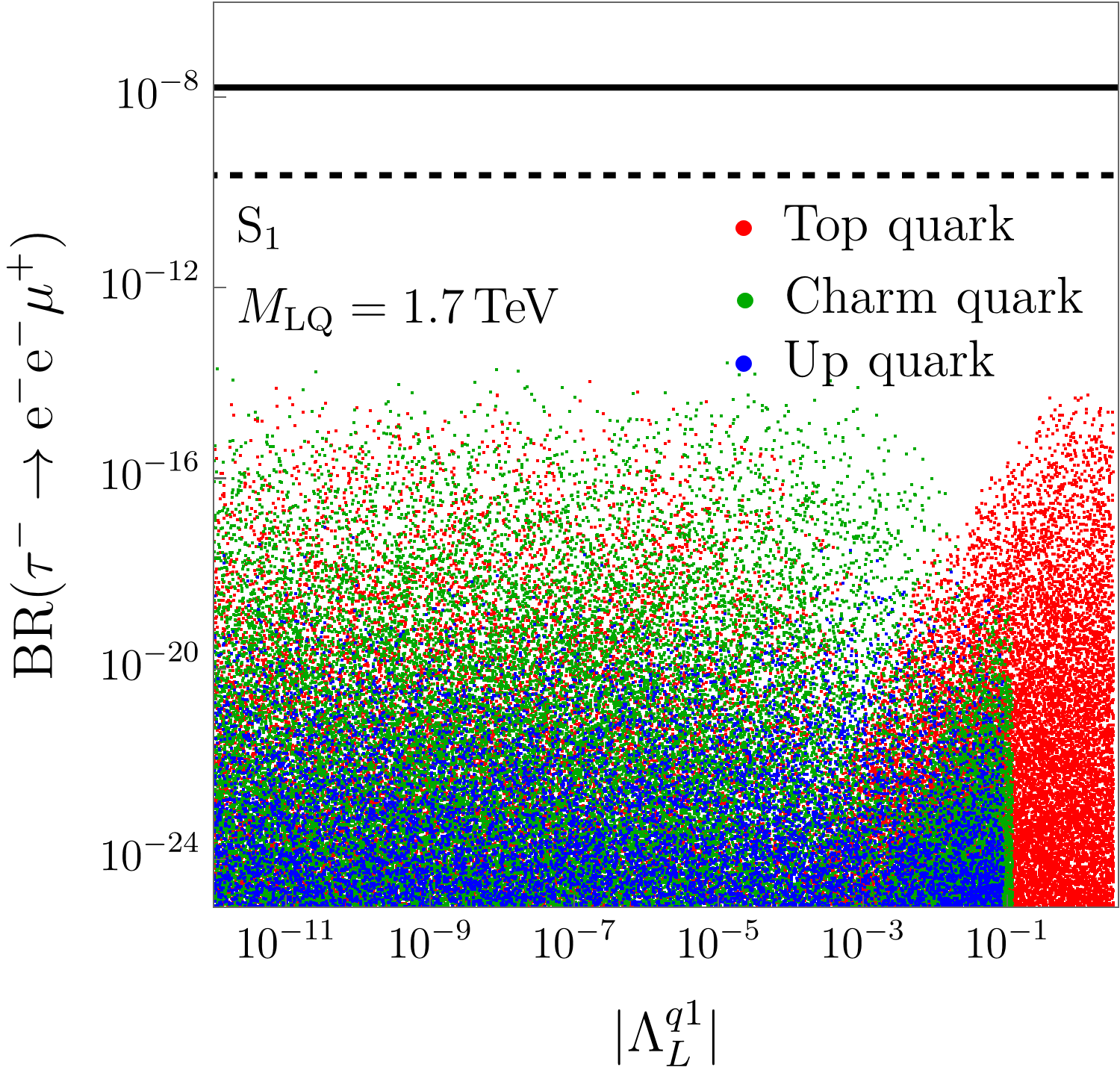}
&
\hspace{0.5cm}
\includegraphics[width=0.455\linewidth]{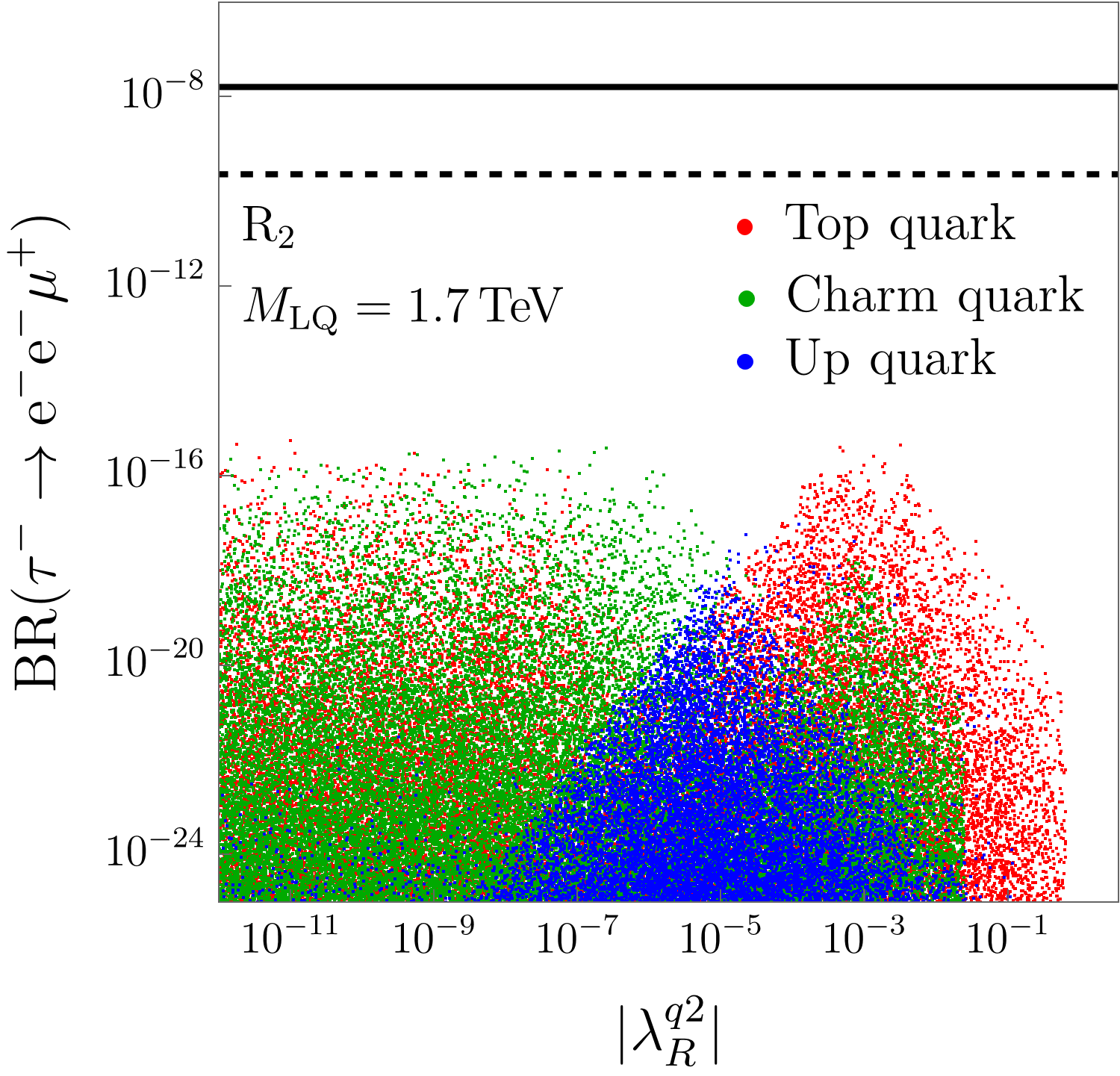}
\\
\includegraphics[width=0.455\linewidth]{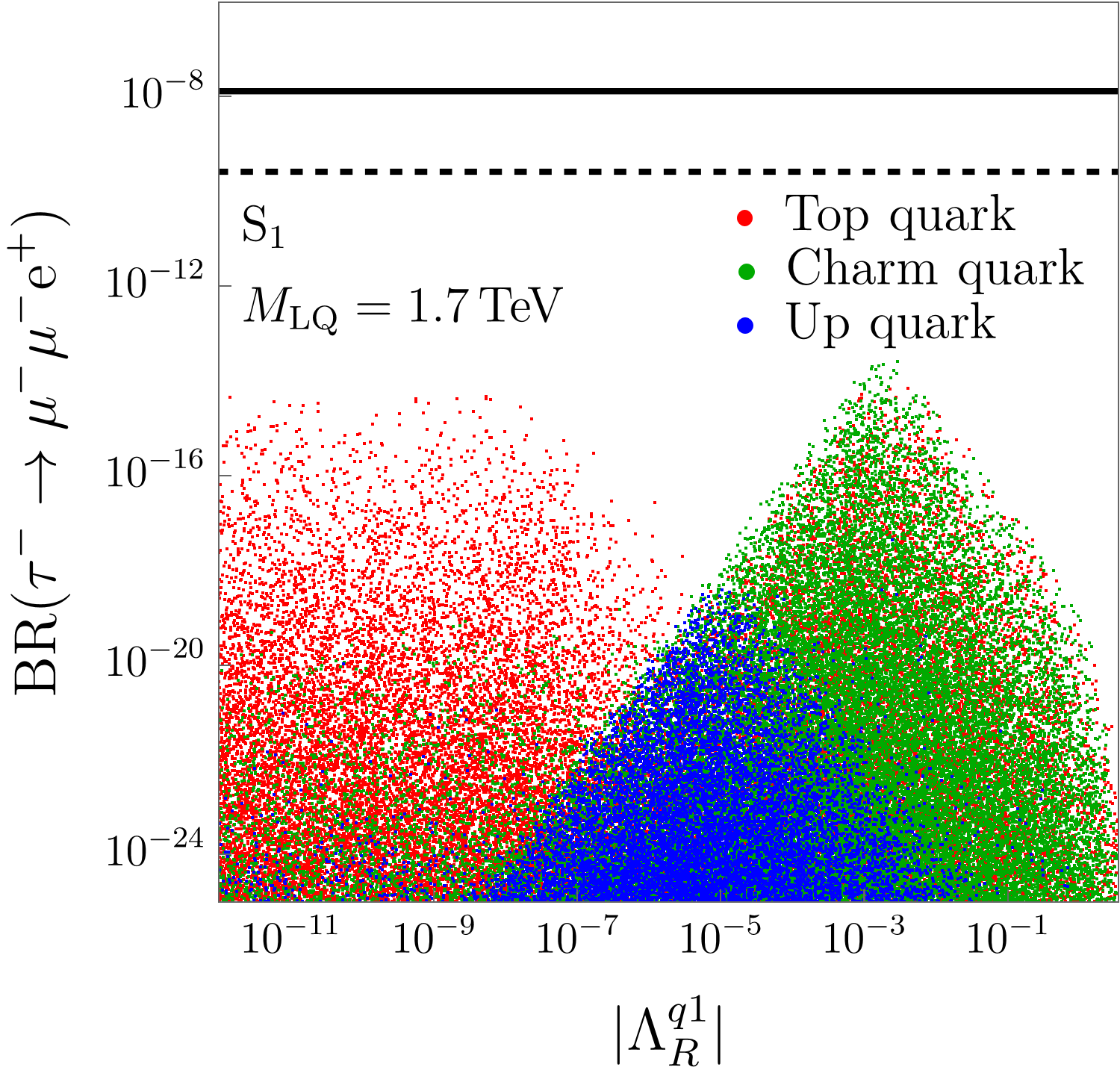}
&
\hspace{0.5cm}
\includegraphics[width=0.455\linewidth]{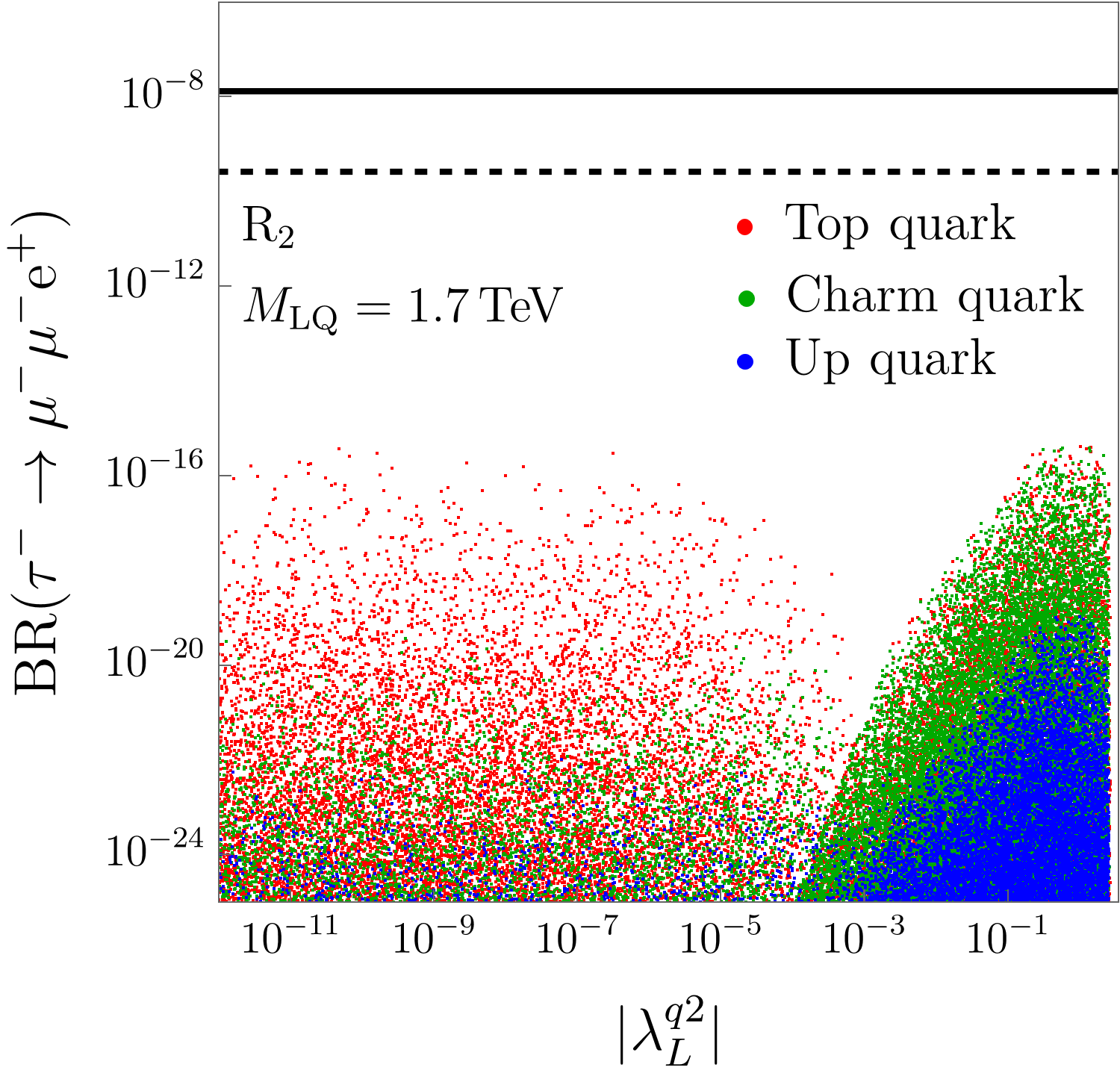}
\end{tabular}
\caption{Branching ratios for the exclusively box-induced tau decays at $M_{\mathrm{LQ}}=1.7~\mathrm{TeV}$. The upper and lower rows show $\tau^-\to e^-e^-\mu^+$ and $\tau^-\to\mu^-\mu^-e^+$, respectively. From left to right the horizontal axes are $|\Lambda_L^{q1}|$ ($S_1$) and $|\lambda_R^{q2}|$ ($R_2$) in the upper row, and $|\Lambda_R^{q1}|$ ($S_1$) and $|\lambda_L^{q2}|$ ($R_2$) in the lower row. Blue, green, and red points denote $q=u,c,t$, respectively. Points satisfy the imposed constraints and the selection $\mathrm{BR}\geq10^{-25}$. Solid and dashed black lines show the Belle~II limit and projected sensitivity specified in Table~\ref{Decay3B}.}\label{BRto2ChS1}
\end{figure}

\begin{figure}[!t]
\centering
\begin{tabular}{cc}
\includegraphics[width=0.455\linewidth]{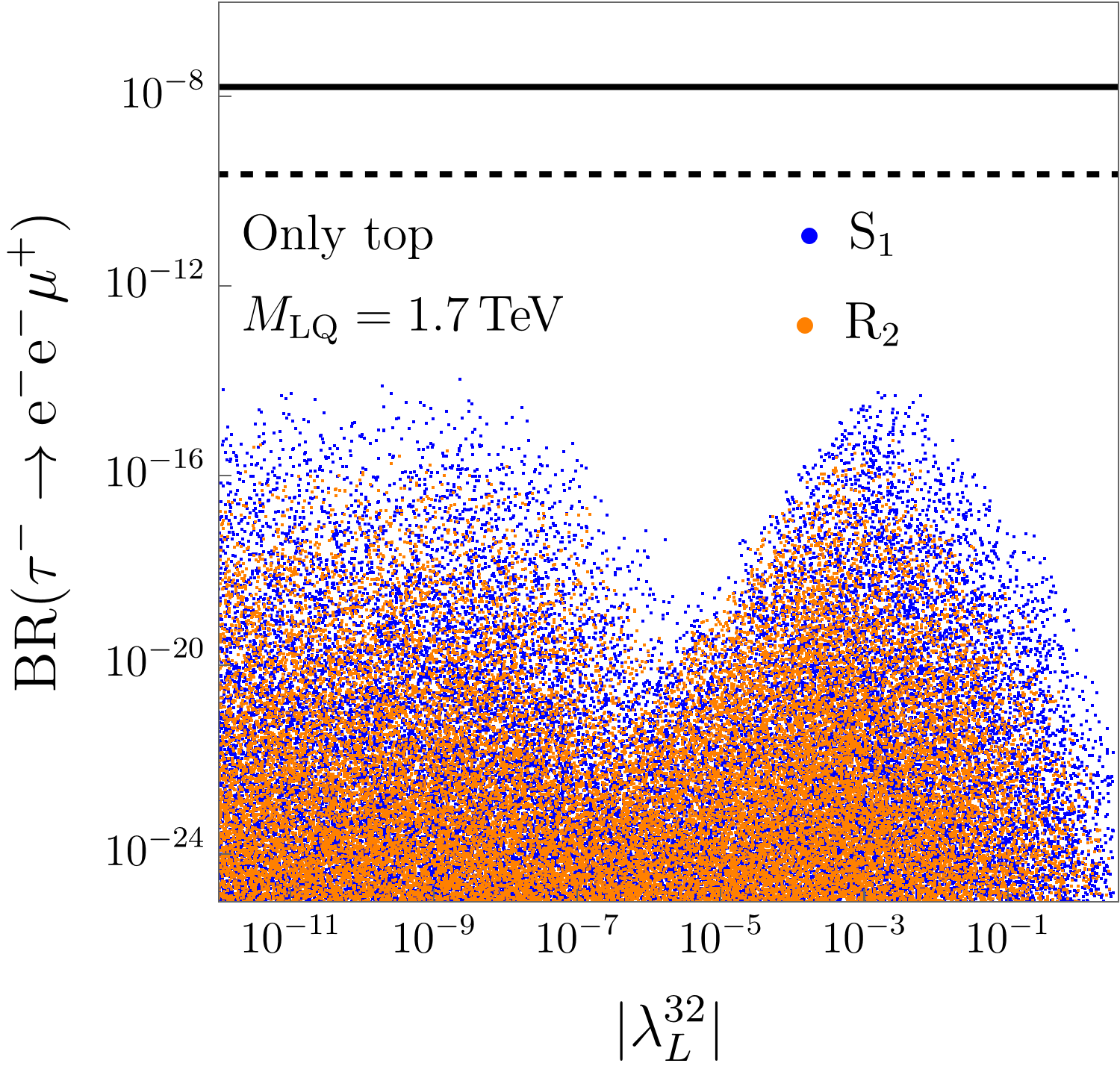}
&
\hspace{0.5cm}
\includegraphics[width=0.455\linewidth]{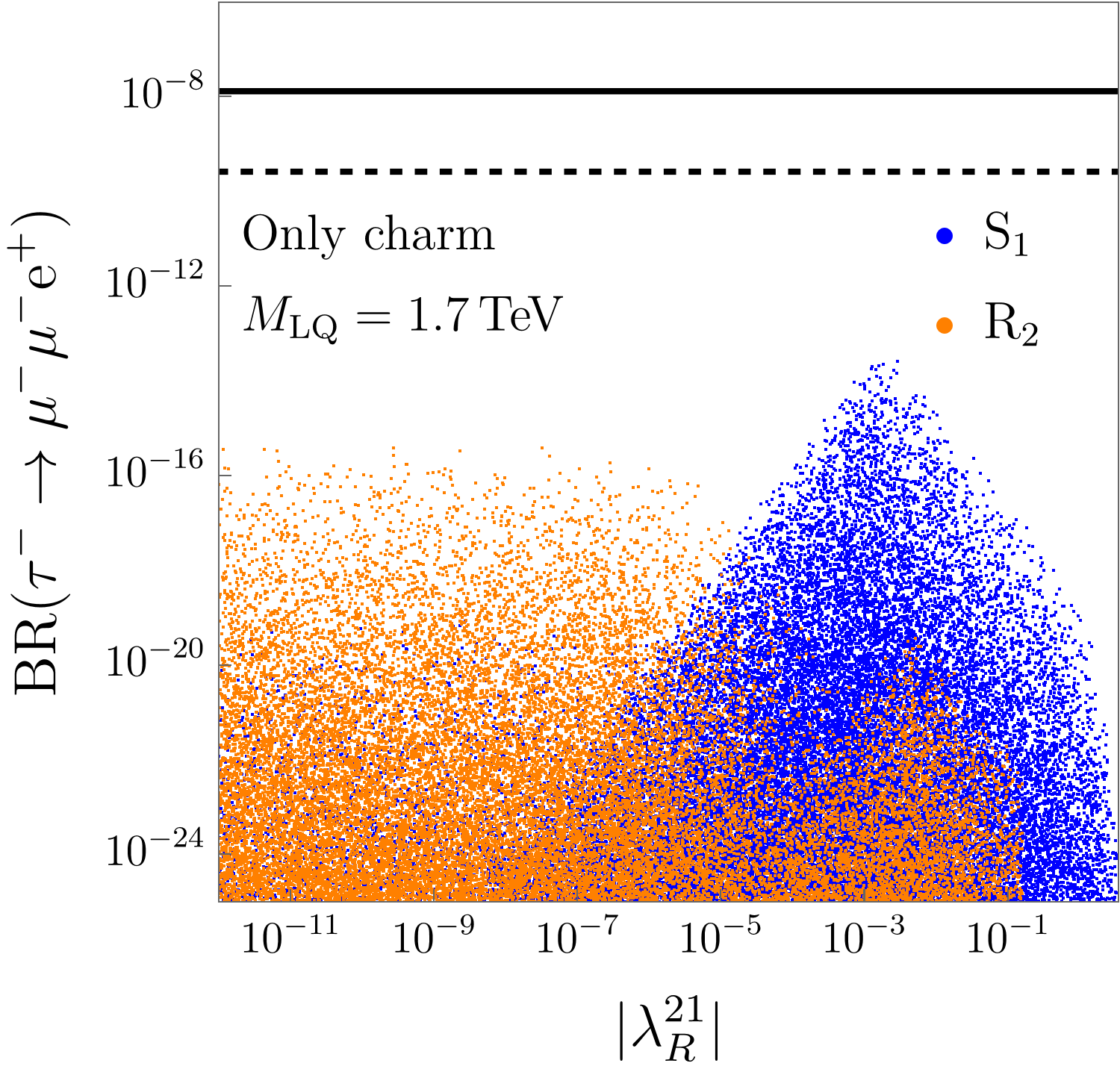}\\
\multicolumn{2}{c}{%
\includegraphics[width=0.455\linewidth]{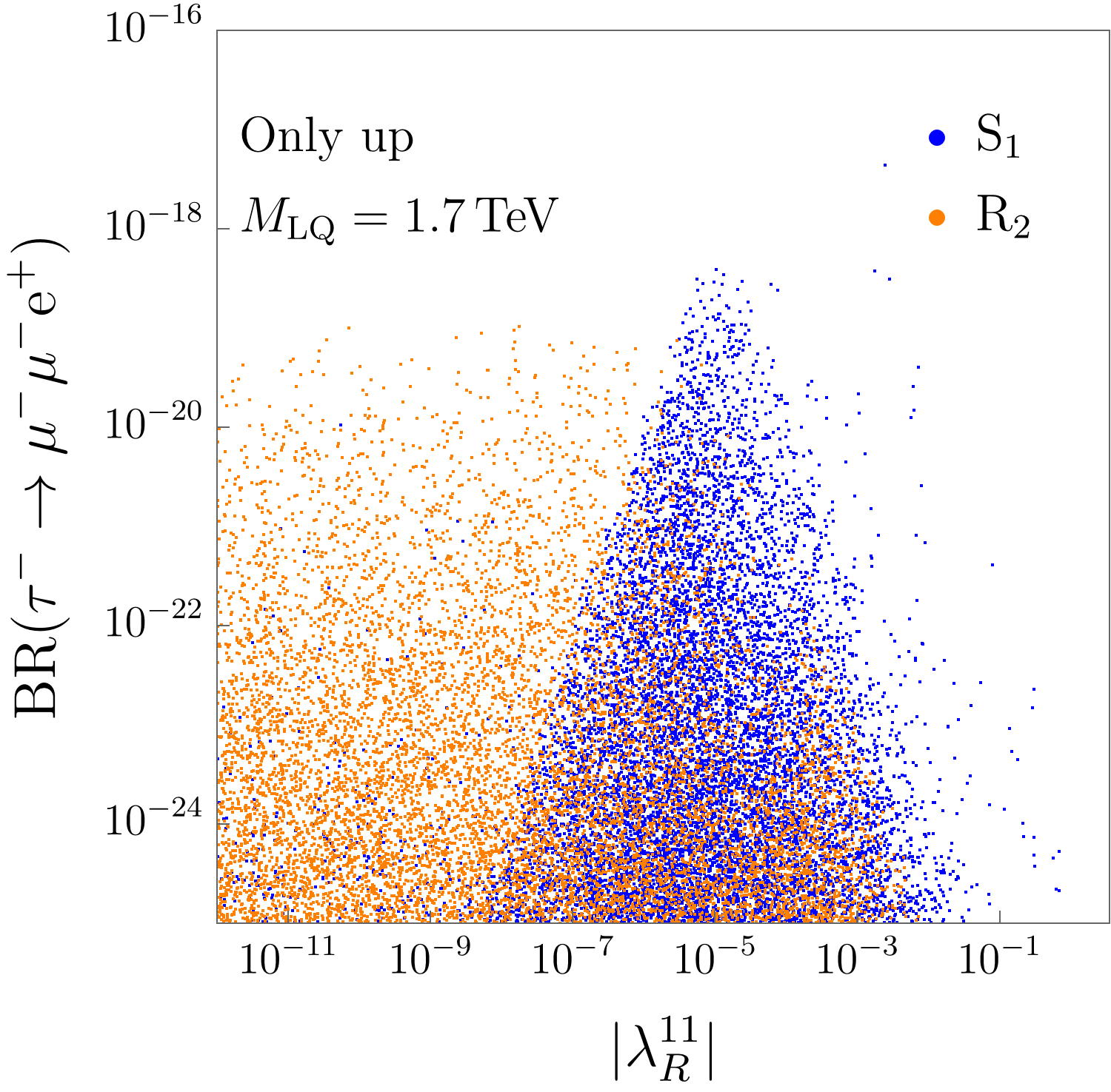}}
\end{tabular}
\caption{Predictions for $S_1$ (blue) and $R_2$ (orange) at $M_{\mathrm{LQ}}=1.7~\mathrm{TeV}$. The upper left panel shows $\tau^-\to e^-e^-\mu^+$ against $|\lambda_L^{32}|$ (top only), the upper right shows $\tau^-\to\mu^-\mu^-e^+$ against $|\lambda_R^{21}|$ (charm only), and the lower panel shows $\tau^-\to\mu^-\mu^-e^+$ against $|\lambda_R^{11}|$ (up only). Here $\lambda$ denotes $\Lambda$ for $S_1$. Points satisfy the imposed constraints and the selection $\mathrm{BR}\geq10^{-25}$. The solid and dashed black lines in the upper panels denote the Belle~II limit and projected sensitivity. Both lie above the displayed range in the lower panel.}
\label{F0vsF2}
\end{figure}

For each of the six combinations of representation and quark flavor, the scan generates $6\times10^6$ candidate six-tuples in advance. It tests them successively until $2\times10^6$ points pass the cuts or the candidate list is exhausted. Each output row records the six accepted couplings in the order $(L_\tau,R_\tau,L_e,R_e,L_\mu,R_\mu)$, where $L_\ell,R_\ell$ denote the two Yukawa couplings to the selected quark. The accompanying Mathematica notebook provides the inputs, observable definitions and cuts. It uses the default random-number generator without setting a seed. For Figs.~\ref{BRto2ChS1} and~\ref{F0vsF2}, the displayed box branching ratio has a lower threshold of $10^{-25}$. This display selection is separate from the cuts used to export couplings.

\begin{figure}[!t]
\centering
\begin{tabular}{cc}
\includegraphics[width=0.455\linewidth]{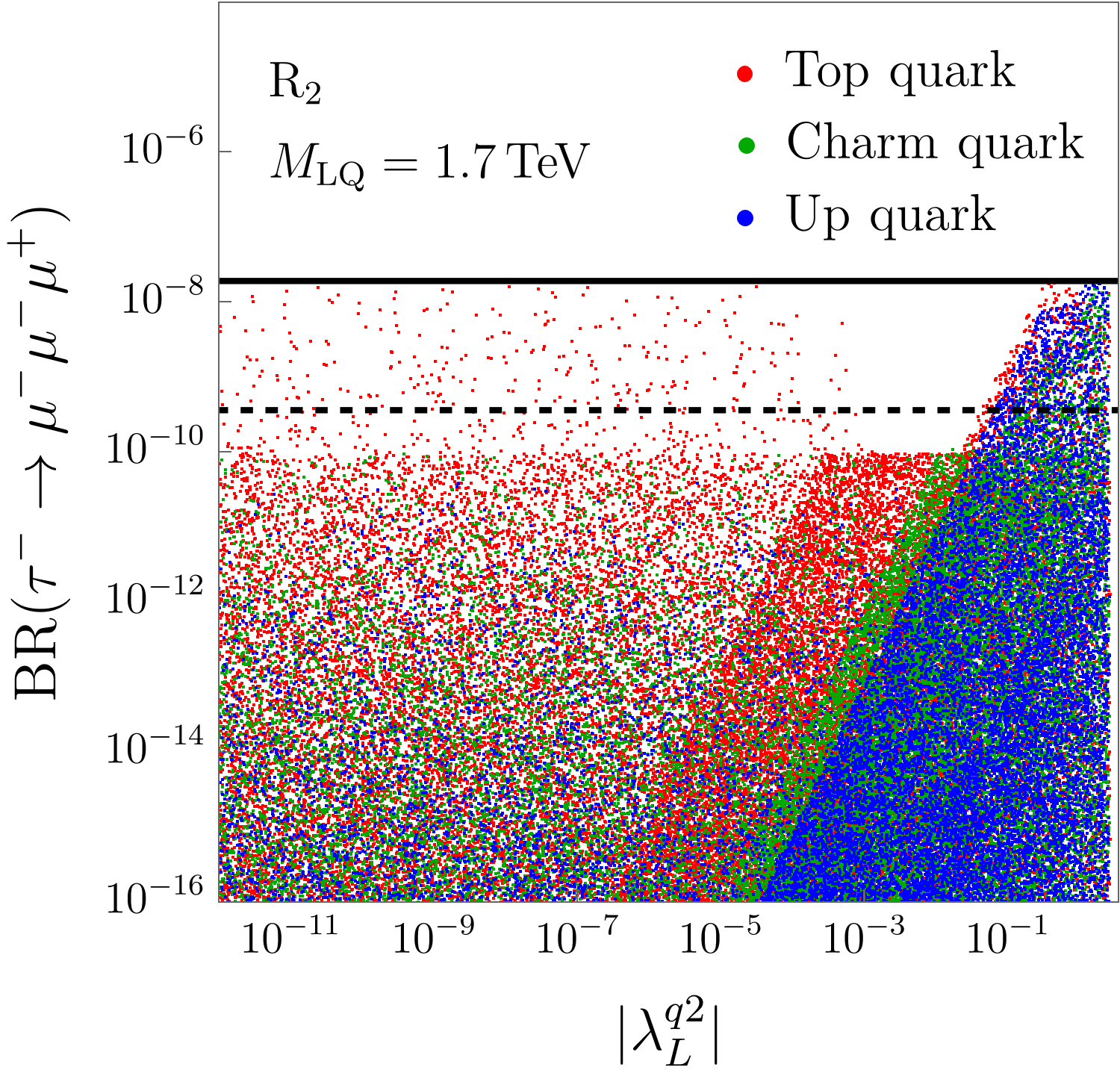}
&
\hspace{0.5cm}
\includegraphics[width=0.455\linewidth]{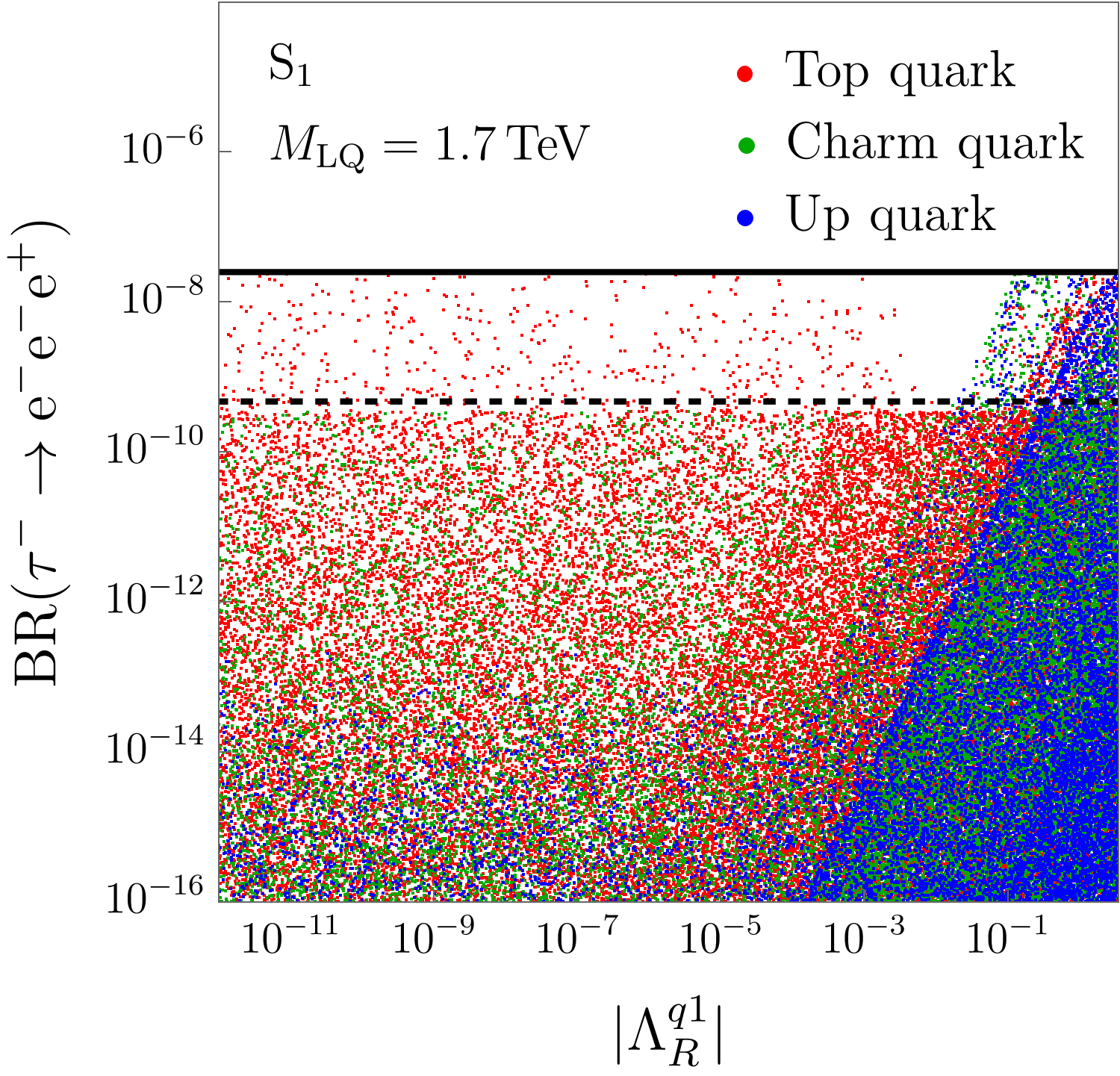}
\\
\includegraphics[width=0.455\linewidth]{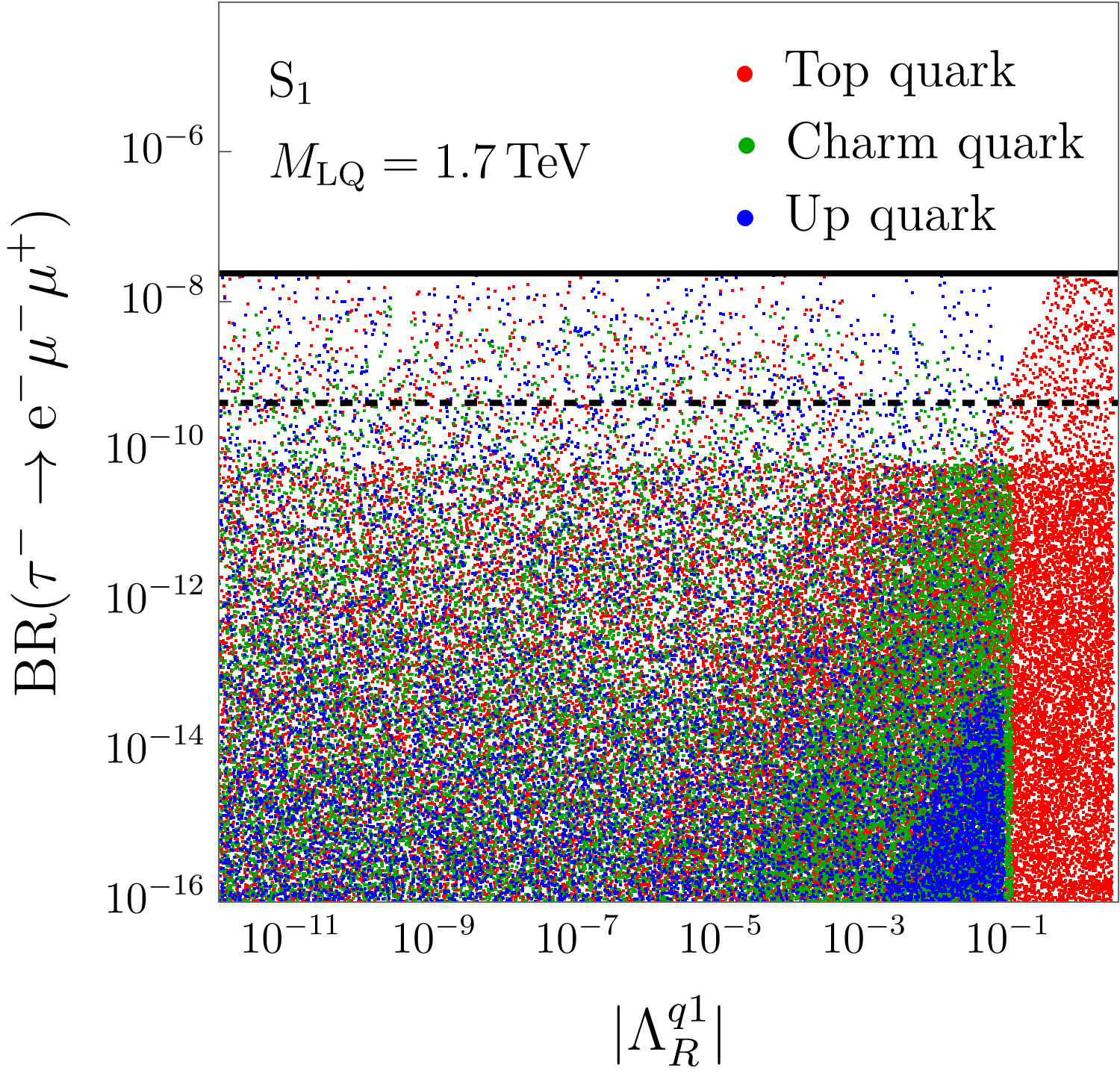}
&
\hspace{0.5cm}
\includegraphics[width=0.455\linewidth]{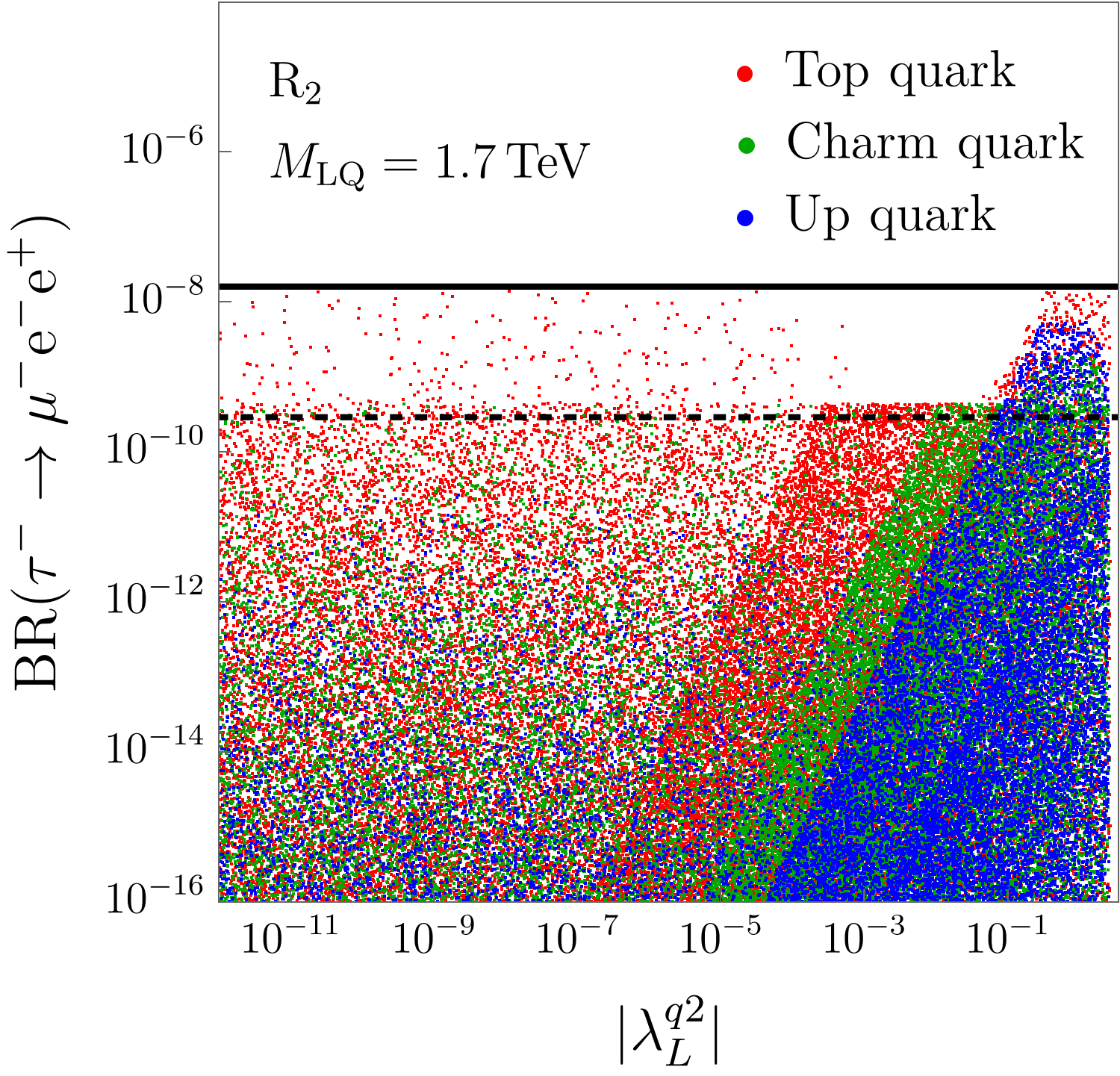}
\end{tabular}
\caption{Representative channels with penguin and box contributions at $M_{\mathrm{LQ}}=1.7~\mathrm{TeV}$. The upper panels show $\tau^-\to\mu^-\mu^-\mu^+$ in $R_2$ against $|\lambda_L^{q2}|$ (left) and $\tau^-\to e^-e^-e^+$ in $S_1$ against $|\Lambda_R^{q1}|$ (right). The lower panels show $\tau^-\to e^-\mu^-\mu^+$ in $S_1$ against $|\Lambda_R^{q1}|$ (left) and $\tau^-\to\mu^-e^-e^+$ in $R_2$ against $|\lambda_L^{q2}|$ (right). Blue, green, and red points denote $q=u,c,t$, respectively. Solid and dashed black lines show the adopted limits and projected sensitivities in Table~\ref{Decay3B}.}
\label{Odecay}
\end{figure}

Using this procedure, we examine the dependence of the branching ratios on the left- and right-handed Yukawa couplings. Figure~\ref{BRto2ChS1} shows the results as functions of the selected couplings for the three scenarios considered. The projections overlap in several coupling ranges, while their vertical spread reflects the simultaneous variation
of the remaining couplings. The point density depends on the sampling prescription and the additional branching-ratio cut and should therefore not be interpreted as a physical probability distribution.

For $S_1$, the largest branching ratios found in the scan for the exclusively box-induced channels are of order $10^{-14}$ in the top-only and charm-only benchmarks and a few $10^{-18}$ to about $10^{-17}$ in the up-only benchmark. The comparison of the branching ratios in the up-only, charm-only, and top-only scenarios is shown in Fig.~\ref{BRto2ChS1}. For $R_2$, the largest plotted values are lower in the top-only and charm-only benchmarks and for $\tau^-\to\mu^-\mu^-e^+$ in the up-only benchmark. Figure~\ref{F0vsF2} shows representative comparisons. These are the largest sampled values, rather than proven global maxima. They remain well below the current experimental limits and projected sensitivities. The difference between the two representations reflects their predictions under the selected constraints and benchmark assumptions and does not imply an experimental preference for either model.

Kaon observables constrain different chiral couplings in the two representations. At tree level, the $R_2^{2/3}$ contributions to
kaon decays into charged leptons involve the $\lambda_R$ couplings, whereas the $S_1$ contribution to $K\to\pi\nu\bar{\nu}$ involves $\Lambda_L$. The resulting restrictions also depend on the corresponding CKM factors. Attributing individual boundaries in the scatter plots to a particular observable requires comparing
scans with that constraint included and removed.

Coherent $\mu$--$e$ conversion primarily restricts the up-only benchmark. In $R_2$, the tree-level contribution retained for the charm-only and top-only benchmarks scales as $|V_{qd}|^4$ in the conversion rate, as shown in Sec.~\ref{mueconversion}.

The suppression of the exclusively box-induced channels is consistent with the restrictions on products of Yukawa couplings. Their amplitudes contain four such couplings, including combinations involving both electron and muon flavors that are also constrained by processes such as $\mu\to e\gamma$, $\mu\to3e$, and coherent $\mu$--$e$ conversion. The resulting predictions reflect the combined effect of the imposed constraints.

For completeness, Fig.~\ref{Odecay} shows representative
predictions from the $S_1$ and $R_2$ models for the four channels
$\tau^-\to\ell_i^-\ell_j^-\ell_j^+$ with
$\ell_i,\ell_j\in\{e,\mu\}$. As already mentioned, these channels receive both penguin and box contributions. We evaluate their branching ratios using Eq.~\eqref{li3lj} for $i=j$ and Eq.~\eqref{lilkljlj} for $i\neq j$. The blue, green, and red points correspond to the up-only, charm-only, and top-only benchmarks, respectively.

Under the same constraints, the branching ratios for the box-only modes remain well below the projected sensitivities throughout the scan, whereas some predictions for the channels in Fig.~\ref{Odecay} approach the current experimental upper
limits. Even in the up-only benchmark, some sampled points
satisfy all imposed constraints and yield branching ratios within
reach of future searches, as shown by the blue points between
the projected sensitivity and the current upper limit in the
corresponding panels.

\section{Conclusions}\label{concl}
In this work, we have studied LFV trilepton tau decays induced by the scalar leptoquarks $S_1$ and $R_2$, with particular emphasis on
$\tau^-\to e^-e^-\mu^+$ and
$\tau^-\to\mu^-\mu^-e^+$.
We have derived analytical expressions for their branching
ratios, which receive contributions exclusively from box
diagrams at one loop. Their phenomenology was investigated in separate up-only, charm-only, and top-only benchmarks, with the Yukawa couplings to the other up-type quark generations set to zero. These choices are assumptions about the flavor structure.

Within the $D_{00}$ approximation and for positive Yukawa couplings, Eq.~\eqref{desLam} translates an experimental branching-ratio limit into an upper bound on the coupling product $\lambda_{RL}^{q3ji^2}$. Replacing that limit with a projected sensitivity gives the corresponding prospective bound.
These bounds illustrate the coupling dependence of the
box-induced channels but do not establish compatibility
with the other phenomenological constraints.

Our numerical analysis assumes real, positive Yukawa couplings and a fixed, illustrative TeV-scale leptoquark mass. We impose a selected set of constraints from $a_\mu$, coherent $\mu$--$e$ conversion at tree level, radiative charged-lepton decays,
$\mu\to3e$, trilepton tau decays, lepton-flavor-violating $Z$ decays, and rare kaon decays. The largest branching ratios found in the scan for the exclusively box-induced tau decays in the top-only and charm-only benchmarks are of order $10^{-14}$ for $S_1$ and $10^{-16}$ for $R_2$, well below current experimental bounds and projected sensitivities. The top-only and charm-only benchmarks yield the largest sampled values of comparable order.
Although the branching-ratio expressions share the same functional form, the parameter points satisfying the imposed constraints yield larger rates for $S_1$ than for $R_2$ within these top-only and charm-only benchmarks.

Tree-level coherent $\mu$--$e$ conversion primarily restricts
the up-only scenario. The CKM-induced down-quark term in $R_2$ also survives for the charm-only and top-only benchmarks. For $S_1$, the light-quark tree-level conversion contribution vanishes
in the charm-only and top-only benchmarks.

The suppression of these channels is consistent with the restrictions on products of Yukawa couplings, including those involving electron and muon flavors. In the single-generation benchmarks, the quartic coupling products entering the box amplitudes can be written as an electron--muon coupling product times a second product involving the tau lepton. Constraints from processes such as $\mu\to e\gamma$, $\mu\to3e$, and coherent $\mu$--$e$ conversion therefore restrict combinations that also enter the two box-induced tau decays. Kaon observables provide complementary restrictions on the couplings.

The exclusively box-induced channels characterize complementary
Yukawa combinations and quantify their suppression under the selected constraints. The other trilepton tau channels, which also
receive penguin contributions, can attain substantially
larger branching ratios under the adopted constraints.
Some sampled points yield rates within the projected
reach of future searches.

Our quantitative conclusions are conditional on the imposed constraints, the up-type-quark loop amplitudes, and the benchmark assumptions. The retained points are not claimed to define globally allowed parameter regions. In particular, constraints from high-energy dilepton tails, leptoquark single production, additional lepton-universality tests, precision measurements of flavor-diagonal electroweak lepton couplings, and loop-induced coherent $\mu$--$e$ conversion have not been included.
Extending the analysis to incorporate these observables, additional hadronic constraints, different leptoquark masses, general coupling signs and complex phases, and simultaneous couplings to multiple quark generations would test the dependence of the results on the assumptions adopted here.
\vspace{-0.5cm}
\section*{Acknowledgments}
J. P. H. D. thanks Consejo Nacional de Humanidades, Ciencia y Tecnología (CONAHCYT) for the financial support during his Ph.D. studies. The work of O. G. M., G.H.T. and R. S. V. has been supported by SNII (Sistema Nacional de Investigadoras e Investigadores, Mexico). R. S. V. also thanks SECIHTI for a postdoctoral grant.
\newpage
\appendix
\section{Amplitude treatment}\label{Amp}

In this appendix, we describe how the amplitudes obtained
in the calculation can be expressed in terms of standard
four-fermion spinor structures. For $S_1$, the interactions involve charge-conjugated quark fields. We use the fermion-flow prescription of Ref.~\cite{DENNER1992467} to treat these vertices
consistently. The appearance of charge-conjugated fields
does not imply physical lepton-number violation:
the interactions and decay channels considered here
conserve total lepton number.

The loop integrals were computed using \texttt{Package-X}
in the limit of vanishing external momenta. This approximation is applied to the loop integrals, while the external momenta are retained in the spinor structures and phase-space integration. For $\tau^-(p)\to\ell_i^-(p_1)\ell_i^-(p_2)\ell_j^+(p_3)$,
the resulting amplitudes contain structures of the form
\begin{equation}
\begin{aligned}
\mathcal{M}_1
&\propto
\left[\bar{u}(p_1)\Gamma_A P_Xu(p)\right]
\left[\bar{u}(p_2)\Gamma^A P_Yv(p_3)\right],
\\
\mathcal{M}_2
&\propto
\left[\bar{u}(p_2)\Gamma_A P_Xu(p)\right]
\left[\bar{u}(p_1)\Gamma^A P_Yv(p_3)\right],
\end{aligned}
\end{equation}
where $\Gamma_A\otimes\Gamma^A$ denotes either
$\mathds{1}\otimes\mathds{1}$ or
$\gamma^\alpha\otimes\gamma_\alpha$,
and $P_X,P_Y\in\{P_L,P_R\}$.

These structures can be related through Fierz transformations, allowing the spinor bilinears to be written in a common ordering. The following identities apply in four dimensions
to external spinors $w_1\equiv w(p_1)$, $w_2\equiv w(p_2)$ (which can be of type $u$ or $v$):
\eq{
&[\bar{w}_1P_{L,R}w_2][\bar{w}_3P_{L,R}w_4]=\frac{1}{2}[\bar{w}_1P_{L,R}w_4][\bar{w}_3P_{L,R}w_2]+\frac{1}{8}[\bar{w}_1\sigma^{\alpha\beta}P_{L,R}w_4][\bar{w}_3\sigma_{\alpha\beta}P_{L,R}w_2],\nonumber\\
&[\bar{w}_1\gamma^\alpha P_{L,R}w_2][\bar{w}_3\gamma_\alpha P_{L,R}w_4]=-[\bar{w}_1\gamma^\alpha P_{L,R}w_4][\bar{w}_3\gamma_\alpha P_{L,R}w_2],\nonumber\\
&[\bar{w}_1P_Lw_2][\bar{w}_3P_Rw_4]=\frac{1}{2}[\bar{w}_1\gamma^\alpha P_Rw_4][\bar{w}_3\gamma_\alpha P_Lw_2],\nonumber\\
&[\bar{w}_1\gamma^\alpha P_Lw_2][\bar{w}_3\gamma_\alpha P_Rw_4]=2[\bar{w}_1P_Rw_4][\bar{w}_3P_Lw_2].
}

The final-state lepton masses are neglected in the phase-space integral. Using the amplitude that already includes the identical-fermion exchange contributions, the differential decay width for the process $\tau^-\to\ell_i^-\ell_i^-\ell_j^+$ is given by
\begin{equation}
\dd[]\Gamma(\tau^-\to\ell_i^-\ell_i^-\ell_j^+)=\frac{\mathscr{F}}{(2\pi)^3 32m_\tau^3}\overline{|\mathcal{M}|^2}\dd[]s_{12}\dd[]s_{23},
\end{equation}
where the bar denotes the sum over final spins
and the average over the initial tau spin. The factor $\mathscr{F}=1/2!$ is the symmetry factor accounting for the two identical fermions in the final state. In the limit of vanishing final-state lepton masses, the integration limits for $s_{12}$ and $s_{23}$ are given by
\begin{equation}
\begin{split}
0\le s_{12}\le m_\tau^2,\quad 0\le s_{23}\le m_\tau^2-s_{12}.
\end{split}
\end{equation}

\section{Details of the box amplitudes}\label{App:BoxDetails}
In this appendix, we collect the form factors entering Eq.~\eqref{BRF0}. The expressions below describe the up-type-quark
contributions associated with $R_2^{5/3}$ and $S_1$. We use $\lambda_{L,R}^{q\ell}$ for the $R_2$ couplings; for $S_1$, the corresponding couplings are denoted by $\Lambda_{L,R}^{q\ell}$.
The lepton index $3$ denotes the tau, while $i,j$ label the final-state flavors in $\tau^-\to\ell_i^-\ell_i^-\ell_j^+$.
\begin{equation}
\begin{split}
S_{LL}&=\frac{N_c}{32\pi^2}\sum_{a,b}\lambda_L^{a3}\lambda_L^{bj}\lambda_R^{ai*}\lambda_R^{bi*}m_{q_a}m_{q_b}D_0^{ab},\\    
S_{LR}&=\frac{N_c}{16\pi^2}\sum_{a,b}\lambda_L^{a3}\lambda_R^{bj}
\left[\lambda_L^{bi*}\lambda_R^{ai*}m_{q_a}m_{q_b}D_0^{ab}-2\lambda_L^{ai*}\lambda_R^{bi*}D_{00}^{ab}\right],\\
V_{LL}&=\frac{N_c}{8\pi^2}\sum_{a,b}\lambda_L^{a3}\lambda_L^{bj}
\lambda_L^{ai*}\lambda_L^{bi*}D_{00}^{ab},\\
V_{LR}&=-\frac{1}{2}S_{LR},\quad T_{LL}=-\frac{1}{4}S_{LL}.
\end{split}
\end{equation}

The overall color factor $N_c=3$ in those expressions arises from the contraction of the internal quark color indices, while $a,b$ label the virtual quark flavors. We define $D_0^{ab}=D_0(x_a,y_b)$ and $D_{00}^{ab}=D_{00}(x_a,y_b)$.

The expressions for $S_{RR}$, $S_{RL}$, $V_{RR}$, $V_{RL}$, and $T_{RR}$ follow from $L\leftrightarrow R$. In particular, $V_{RL}=-S_{RL}/2$ and $T_{RR}=-S_{RR}/4$. The tensor and mixed-chirality vector contributions are incorporated through these relations before obtaining Eq.~\eqref{BRF0}; they are not discarded.

In the zero-external-momentum limit, the relevant
Passarino--Veltman functions are~\cite{PASSARINO1979151}
\eq{
D_0(x_a,y_b)&=-\frac{1}{M_{\rm{LQ}}^4}\left[\frac{1}{(x_a-1)(y_b-1)}+\frac{x_a\log x_a}{(x_a-1)^2(x_a-y_b)}+\frac{y_b\log y_b}{(y_b-1)^2(y_b-x_a)}\right],\nonumber\\
D_{00}(x_a,y_b)&=-\frac{1}{4M_{\rm{LQ}}^2}\left[\frac{1}{(x_a-1)(y_b-1)}+\frac{x_a^2\log x_a}{(x_a-1)^2(x_a-y_b)}+\frac{y_b^2\log y_b}{(y_b-1)^2(y_b-x_a)}\right],
}
where $x_a=m_{q_a}^2/M_{\mathrm{LQ}}^2$ and $y_b=m_{q_b}^2/M_{\mathrm{LQ}}^2$. For equal internal quark masses, as in the single-generation benchmarks, these functions reduce to~\cite{PhysRevD.108.095027}
\begin{equation}\label{PassRed}
\begin{split}
D_0(x_q)&\equiv D_0^q=\frac{2-2x_q+(x_q+1)\log x_q}{M_{\rm{LQ}}^4(x_q-1)^3},\\
D_{00}(x_q)&\equiv D_{00}^q=\frac{1-x_q^2+2x_q\log x_q}{4M_{\rm{LQ}}^2(x_q-1)^3}.
\end{split}
\end{equation}

In the light-quark limit,
\begin{equation}
\lim_{x_q\to0}D_{00}^q
=-\frac{1}{4M_{\mathrm{LQ}}^2},
\qquad
\lim_{x_q\to0}m_q^2D_0^q=0.
\end{equation}

Although $D_0^q$ itself contains a logarithmic divergence in this limit, its mass-weighted contribution vanishes. For $x_q\ll1$, $m_q^2D_0^q/D_{00}^q=\mathcal{O}(x_q\ln x_q)$. This loop-function hierarchy motivates the $D_{00}$ approximation for suitable coupling configurations, but does not justify dropping the mass-weighted terms uniformly over arbitrary Yukawa hierarchies.

Restricting the calculation to one quark generation, $q=1,2,3$, and real Yukawa couplings, Eq.~\eqref{BRF0} becomes
\eq{\label{BRF0num}
\mathrm{BR}&=\frac{m_\tau^5}{512\pi^3\Gamma_\tau}
\left(\frac{N_c}{32\sqrt{6}\pi^2}\right)^2\nonumber\\
&\times\Bigg\{8(D_{00}^q)^{2}\left(2\left[\lambda_L^{q3}\lambda_L^{qj}(\lambda_L^{qi})^2\right]^2
+2\left[\lambda_R^{q3}\lambda_R^{qj}(\lambda_R^{qi})^2\right]^2+\left[\lambda_L^{qi}\lambda_L^{qj}\lambda_R^{q3}\lambda_R^{qi}\right]^2+\left[\lambda_L^{q3}\lambda_L^{qi}\lambda_R^{qi}\lambda_R^{qj}\right]^2
\right)\nonumber\\
&+(m_q^2D_0^q)^2\left(\left[\lambda_L^{q3}\lambda_L^{qj}(\lambda_R^{qi})^2\right]^2+\left[\lambda_R^{q3}\lambda_R^{qj}(\lambda_L^{qi})^2\right]^2+2\left[\lambda_L^{qi}\lambda_L^{qj}\lambda_R^{q3}\lambda_R^{qi}\right]^2
+2\left[\lambda_L^{q3}\lambda_L^{qi}\lambda_R^{qi}\lambda_R^{qj}\right]^2\right)\nonumber\\
&-8m_q^2D_0^qD_{00}^q\left(\left[\lambda_L^{qi}\lambda_L^{qj}\lambda_R^{q3}\lambda_R^{qi}\right]^2+\left[\lambda_L^{q3}\lambda_L^{qi}\lambda_R^{qi}\lambda_R^{qj}\right]^2\right)\Bigg\}.
}

Neglecting $m_q^2D_0^q$ reproduces Eq.~\eqref{BReq}. For complex couplings, the branching ratio must instead be evaluated using the complex form factors in Eq.~\eqref{BRF0}, retaining the appropriate
conjugations and interference terms.

\section{Branching ratios for
\texorpdfstring{$\ell_i^-\to\ell_j^-\ell_k^-\ell_m^+$}
{li- -> lj- lk- lm+}}
\label{Oproc}

\subsection{Decays with identical final-state flavors}
Within the approximation specified in Sec.~\ref{threebodyconstraints}, the branching ratio is
\eq{\label{li3lj}
\text{BR}(\ell_i\to3\ell_j)&=\frac{m_i^5}{512\pi^3\Gamma_i}
\Bigg\{Q_{\mathrm{em}}^4\biggl[|A_{1L}|^2+|A_{1R}|^2-4\text{Re}\left(A_{1L}A_{2R}^*+A_{1R}A_{2L}^*\right)\nonumber\\
&+\left(|A_{2L}|^2+|A_{2R}|^2\right)\left(\frac{16}{3}\ln\left(\frac{m_i}{m_j}\right)-\frac{22}{3}\right)\biggr]\nonumber\\
&+\frac{1}{6}\left(|S_{LL}|^2+|S_{RR}|^2+|V_{LL}|^2+|V_{RR}|^2\right)+\frac{1}{12}\left(|S_{LR}|^2+|S_{RL}|^2\right)\nonumber\\
&+\frac{Q_{\rm{em}}^4}{3}\left[2\left(\left|g_L^\ell F_{1L}\right|^2+\left|g_R^\ell F_{1R}\right|^2\right)+\left(\left|g_R^\ell F_{1L}\right|^2+\left|g_L^\ell F_{1R}\right|^2\right)\right]\nonumber\\
&+\frac{2Q_{\rm em}^2}{3}\text{Re}\left(A_{1L}V_{LL}^*+A_{1R}V_{RR}^*-2A_{2R}V_{LL}^*-2A_{2L}V_{RR}^*\right)\\
&-\frac{Q_{\rm em}^2}{3}\text{Re}\left(A_{1L}S_{LR}^*+A_{1R}S_{RL}^*-2A_{2R}S_{LR}^*-2A_{2L}S_{RL}^*\right)\nonumber\\
&+\frac{2Q_{\rm em}^2}{3}\text{Re}\left(V_{LL}F_{1L}^*g_L^\ell+V_{RR}F_{1R}^*g_R^\ell\right)\nonumber\\
&-\frac{Q_{\rm em}^2}{3}\text{Re}\left( S_{LR}F_{1L}^*g_R^\ell+S_{RL}F_{1R}^*g_L^\ell\right)\nonumber\\
&+\frac{2Q_{\rm em}^4}{3}\text{Re}\left[2\left(A_{1L}F_{1L}^*g_L^\ell-A_{2R}F_{1L}^*g_R^\ell\right)+A_{1L}F_{1L}^*g_R^\ell-4A_{2R}F_{1L}^*g_L^\ell\right]\nonumber\\
&+\frac{2Q_{\rm em}^4}{3}\text{Re}\left[2\left(A_{1R}F_{1R}^*g_R^\ell-A_{2L}F_{1R}^*g_L^\ell\right)
+A_{1R}F_{1R}^*g_L^\ell-4A_{2L}F_{1R}^*g_R^\ell\right]
\Bigg\}\nonumber.
}
The photon form factors $A_{1X}$ and $A_{2X}$
describe vector and dipole contributions, respectively,
while $F_{1X}$ describes the retained vector $Z$ contribution.
The box form factors are denoted by $S_{XY}$ and $V_{XY}$,
with $X,Y=L,R$.
In the reduced box expression, the relations
$V_{LR}=-S_{LR}/2$ and $T_{LL}=-S_{LL}/4$,
together with their $L\leftrightarrow R$ counterparts,
have been used.
\eq{\nonumber
A_{1L}&=\frac{N_c}{288\pi^2M_{\rm LQ}^2}\sum_q\lambda_L^{qj*}\lambda_L^{qi}\left\{Q_{\rm LQ}\Big[1-3x_qF_2(x_q)\Big]
-3Q_q\Big[F_1(x_q)+F_3(x_q)\Big]\right\}\\[0.25cm] 
A_{2L}&=\frac{N_c}{32\pi^2 M_{\mathrm{LQ}}^2}\sum_q\Bigg\{Q_{\mathrm{LQ}}
\left[\frac{m_q}{m_i}\lambda_L^{qi}\lambda_R^{qj*}F_4(x_q)+\lambda_R^{qi}\lambda_R^{qj*}F_2(x_q)\right]\label{FacFA}\\
&\hspace{3.5cm}+Q_q\left[\frac{m_q}{m_i}\lambda_L^{qi}\lambda_R^{qj*}F_3(x_q)+\lambda_R^{qi}\lambda_R^{qj*}F_1(x_q)\right]\Bigg\}\nonumber\\[0.25cm]
F_{1L}&=-\frac{N_c}
{32\pi^2 c_W^2 s_W^2M_Z^2M_{\mathrm{LQ}}^2}\sum_q\left(g_L^q-g_R^q\right)m_q^2\lambda_L^{qj*}\lambda_L^{qi}\left[F_3(x_q)+F_4(x_q)\right]\label{FacFZ}\\[0.25cm]
S_{LL}&=\frac{N_c}{32\pi^2}\sum_{a,b}\lambda_{R}^{aj*} \lambda_{R}^{bj*}\lambda_{L}^{ai}\lambda_{L}^{bj}m_{q_a}m_{q_b}D_0^{ab},\nonumber\\
S_{LR}&=\frac{N_c}{16\pi^2}\sum_{a,b}\lambda_L^{ai}\lambda_R^{bj}
\left(\lambda_L^{bj*}\lambda_R^{aj*}m_{q_a}m_{q_b}D_0^{ab}
-2\lambda_L^{aj*}\lambda_R^{bj*}D_{00}^{ab}\right),\label{FacFB3l}\\
V_{LL}&=\frac{N_c}{8\pi^2}\sum_{a,b}\lambda_L^{aj*}\lambda_L^{bj*} \lambda_L^{ai}\lambda_L^{bj}D_{00}^{ab}\nonumber.
}
Here, $s_W=\sin\theta_W$ and $c_W=\cos\theta_W$.
The opposite-chirality form factors are obtained by interchanging
$L$ and $R$; for $F_{1R}$, this interchange also applies to
$g_L^q$ and $g_R^q$.\\
The quantities $g_X^f$, with $f=\ell,q$, denote the
dimensionless chiral couplings of a fermion $f$ to the $Z$ boson, defined by
\begin{equation}
g_{L,R}^{f}=T_3^{f_{L,R}}-Q_fs_W^2.
\end{equation}
For charged leptons and up-type quarks, these couplings take the values
\begin{equation}
g_L^\ell=-\frac12+s_W^2,\quad g_R^\ell=s_W^2,\quad g_L^u=\frac12-\frac23s_W^2,\quad g_R^u=-\frac23s_W^2,
\end{equation}
which apply to the $R_2^{5/3}$ case. For $S_1$, the Yukawa couplings are replaced by $\Lambda_{L,R}$,
the charge pair $(Q_q,Q_{\mathrm{LQ}})=(-2/3,1/3)$ is used, and the $Z$ couplings
of the charge-conjugated quark fields are used.
For charge-conjugated up-type quarks, these couplings are
\begin{equation}
g_L^{u^c}=-g_R^u=\frac23s_W^2,
\qquad
g_R^{u^c}=-g_L^u=-\frac12+\frac23s_W^2,
\end{equation}
reflecting the reversal of the gauge charges and the interchange
of the left- and right-handed components under charge conjugation.

With the dipole conventions used here,
$A_{2L}=C_L^{ij}/m_i$ and $A_{2R}=C_R^{ij}/m_i$ in the limit $m_j\to0$. This provides a consistency relation between the
radiative and three-body decay expressions.

\subsection{Decays with distinct final-state flavors}
In this subsection, we give the branching ratio for $\ell_i^-\to\ell_k^-\ell_j^-\ell_j^+$ with $k\neq j$, including the channels
$\tau^-\to e^-\mu^-\mu^+$ and $\tau^-\to\mu^-e^-e^+$. Here, $i$ labels the decaying lepton, whereas $k$ and $j$ label distinct final-state flavors. We use the spinor ordering $[\bar{u}_k(p_1)\Gamma u_i(p)][\bar{u}_j(p_2)\Gamma'v_j(p_3)]$. Unlike the channels with two identical negatively charged leptons, these decays do not require a $1/2!$ phase-space symmetry factor.

We retain photon-penguin, vector $Z$-penguin, and box contributions. Final-state masses are neglected except for $m_j$
in the collinear logarithm associated with the virtual photon.
Dipole $Z$-penguin contributions and additional mass-suppressed terms are omitted. Within these approximations,
\eq{\nonumber
\text{BR}(\ell_i^-\to\ell_k^-\ell_j^-\ell_j^+)&=\frac{m_i^5}{512\pi^3\Gamma_i}
\Biggl\{Q_{\mathrm{em}}^4\Biggl[\frac{2}{3}
\left(|A_{1L}|^2+|A_{1R}|^2\right)+
\left(|A_{2L}|^2+|A_{2R}|^2\right)\left[\frac{16}{3}\ln\left(\frac{m_i}{m_j}\right)-8\right]\\
&-\frac{8}{3}\text{Re}\left(A_{1L}A_{2R}^{*}+A_{1R}A_{2L}^{*} \right)\Biggr]+
\frac{1}{12}\left(|S_{LL}|^2+|S_{LR}|^2+|S_{RL}|^2+|S_{RR}|^2\right)\nonumber\\
&+\frac{1}{3}\left(|V_{LL}|^2+|V_{LR}|^2+|V_{RL}|^2+|V_{RR}|^2\right)
+4\left(|T_{LL}|^2+|T_{RR}|^2\right)\nonumber\\
&+\frac{Q_{\rm{em}}^4}{3}\left(
\left|F_{1L}g_L^\ell\right|^2
+\left|F_{1R}g_R^\ell\right|^2
+\left|F_{1L}g_R^\ell\right|^2
+\left|F_{1R}g_L^\ell\right|^2
\right)\nonumber\\
&+\frac{2Q_{\rm em}^2}{3}\text{Re}\left(
A_{1L}V_{LL}^*+A_{1R}V_{RR}^*
-2A_{2R}V_{LL}^*-2A_{2L}V_{RR}^*\right)\label{lilkljlj}\\
&+\frac{2Q_{\rm em}^2}{3}\text{Re}\left(
A_{1L}V_{LR}^*+A_{1R}V_{RL}^*
-2A_{2R}V_{LR}^*-2A_{2L}V_{RL}^*\right)\nonumber\\
&+\frac{2Q_{\rm em}^2}{3}\text{Re}\left(
V_{LL}F_{1L}^*g_L^\ell+V_{RR}F_{1R}^*g_R^\ell\right)\nonumber\\
&+\frac{2Q_{\rm em}^2}{3}\text{Re}\left(
V_{LR}F_{1L}^*g_R^\ell+V_{RL}F_{1R}^*g_L^\ell\right)\nonumber\\
&+\frac{2Q_{\rm em}^4}{3}\text{Re}\left(
A_{1L}F_{1L}^*g_L^\ell+A_{1L}F_{1L}^*g_R^\ell
-2A_{2R}F_{1L}^*g_L^\ell-2A_{2R}F_{1L}^*g_R^\ell\right)\nonumber\\
&+\frac{2Q_{\rm em}^4}{3}\text{Re}\left(
A_{1R}F_{1R}^*g_R^\ell+A_{1R}F_{1R}^*g_L^\ell
-2A_{2L}F_{1R}^*g_R^\ell-2A_{2L}F_{1R}^*g_L^\ell\right)
\Biggr\}\nonumber.
}
The form factors $A_{1X}$, $A_{2X}$,
and $F_{1X}$ are obtained from
Eqs.~\eqref{FacFA} and~\eqref{FacFZ}
by replacing the outgoing-lepton index $j$
with $k$.
This replacement applies to the flavor indices
in those form factors; the mass $m_j$ in
Eq.~\eqref{lilkljlj} remains the mass of the
lepton pair produced by the virtual photon.\\
The up-type-quark box form factors for these
channels are
\begin{equation}
\begin{split}
S_{LL}&=\frac{N_c}{32\pi^2}\sum_{a,b}\left(2\lambda_R^{ak*}\lambda_R^{bj*}-\lambda_R^{aj*}\lambda_R^{bk*}\right)m_a m_b
\lambda_L^{ai}\lambda_L^{bj}D_0^{ab},\\
S_{LR}&=\frac{N_c}{16\pi^2}\sum_{a,b}\lambda_L^{ai}\lambda_R^{bj}
\left(m_am_b\lambda_R^{ak*}\lambda_L^{bj*}D_0^{ab}-2\lambda_L^{aj*}\lambda_R^{bk*}D_{00}^{ab}\right),\\
V_{LL}&=\frac{N_c}{16\pi^2}\sum_{a,b}\left(\lambda_L^{ak*}\lambda_L^{bj*}+\lambda_L^{aj*}\lambda_L^{bk*}\right)\lambda_L^{ai}\lambda_L^{bj}D_{00}^{ab},\\
V_{LR}&=\frac{N_c}{32\pi^2}\sum_{a,b}\lambda_L^{ai}\lambda_R^{bj}
\left(2\lambda_L^{ak*}\lambda_R^{bj*}D_{00}^{ab}-m_a m_b\lambda_R^{aj*}\lambda_L^{bk*}D_0^{ab}\right),\\
T_{LL}&=-\frac{N_c}{128\pi^2}\sum_{a,b}m_a m_b\lambda_L^{ai}\lambda_L^{bj} \lambda_R^{aj*}\lambda_R^{bk*}D_0^{ab}.    
\end{split}    
\end{equation}
The remaining form factors follow from
$L\leftrightarrow R$. For $k\neq j$, the relations
$T_{LL}=-S_{LL}/4$ and $V_{LR}=-S_{LR}/2$
do not hold in general, so the tensor and
mixed-chirality vector terms are retained explicitly
in Eq.~\eqref{lilkljlj}.
\section{Branching ratios for \texorpdfstring{$Z\to\ell_i^\pm\ell_j^\mp$}{Z -> li+- lj-+}}\label{BRZ}
\begin{table}[!b]
\caption{Observed upper limits at 95\% C.L. on the branching
ratios of lepton-flavor-violating $Z$ decays. Both charge
configurations are included.}\label{TBZLFV}
\begin{ruledtabular}
\begin{tabular}{cc}
Process&Current limit 95$\%$ C.L.\\ \hline
$\text{BR}(Z\to e\mu)$& $1.9\times10^{-7}$~\cite{CMS2025ZLFV}\\
$\text{BR}(Z\to e\tau)$&$5.0\times10^{-6}$~\cite{PhysRevLett.127.271801}\\ 
$\text{BR}(Z\to\mu\tau)$&$6.5\times10^{-6}$~\cite{PhysRevLett.127.271801}\\
\end{tabular}
\end{ruledtabular}
\end{table}
In scalar leptoquark models, the decay $Z\to\ell_i^+\ell_j^-$ is described by the Feynman diagrams shown in Figs.~\ref{DFPin1} and~\ref{DFPin2}, adapted to this process. Table~\ref{TBZLFV} lists the experimental upper limits. Retaining only the vector form factors~\cite{Crivellin2021,Arnan2019}, we obtain the following amplitude
\begin{equation}\label{ampZ}
\mathcal{M}(Z\to\ell_i^+\ell_j^-)=\frac{Q_{\text{em}}}{s_Wc_W}\bar{u}_j(p_1)\gamma_\alpha\left(V_LP_L+V_RP_R\right)v_i(p_2)\epsilon^\alpha_Z(p),
\end{equation}
where the form factors $V_{L,R}$ are obtained by neglecting the external lepton masses and treating the contributions from the top quark and the light quarks separately. The dependence on $x_t=m_t^2/M_{\mathrm{LQ}}^2$ is retained exactly in the top contribution, whereas the light-quark contributions are evaluated using the hierarchy $m_q^2\ll M_Z^2\ll M_{\mathrm{LQ}}^2$. Terms are retained through first order in $x_Z=M_Z^2/M_{\mathrm{LQ}}^2$, including the associated logarithmic terms. The resulting form factors are given by
\begin{equation}
\begin{split}
V_L&=\frac{N_c}{16\pi^2}\lambda_L^{tj*}\lambda_L^{ti}\left[(g_L^t-g_R^t)\frac{x_t(x_t-1-\ln x_t)}{(x_t-1)^2}+\frac{x_Z}{6}T(x_t)\right]\\&+\frac{N_c}{48\pi^2}x_Z\sum_{q\neq t }\lambda_L^{qj*}\lambda_L^{qi}\left[\frac{g_L^\ell}{6}-g_R^q\left(\ln x_Z-i\pi-\frac16\right)\right],\\
V_R&=V_L(L\leftrightarrow R).
\end{split}
\end{equation}
The function $T(x_t)$ is given by
\begin{equation}
\begin{split}
T(x_t)=3g_L^tF_1(x_t)+g_R^t\left(x_tF_2(x_t)+F_3(x_t)-2F_1(x_t)-\frac13\right)+g_L^\ell\left(\frac13-x_tF_2(x_t)\right).    
\end{split}
\end{equation}
From Eq.~\eqref{ampZ}, we obtain the decay width:
\begin{equation}
\begin{split}
\Gamma(Z\to\ell_i^+\ell_j^-)&=\frac{Q_{\text{em}}^2\lambda_Z^{1/2}}{24\pi c_W^2 s_W^2 M_Z}\Bigg\{
\left[1-\frac{(m_j^2+m_i^2)}{2M_Z^2}-\frac{(m_j^2-m_i^2)^2}{2M_Z^4}\right]\left(|V_L|^2+|V_R|^2\right)\\
&+6\frac{m_jm_i}{M_Z^2}\text{Re}\left(V_L^*V_R\right)
\Bigg\},\\
\lambda_Z&=\left[M_Z^2-(m_j+m_i)^2\right]\left[M_Z^2-(m_j-m_i)^2\right].
\end{split}
\end{equation}
For the charge-conjugate final state, interchange the flavor labels $i$ and $j$ with the negatively charged lepton represented by the $u$ spinor. For the real Yukawa couplings considered here,
\begin{equation}
\Gamma(Z\to\ell_i^+\ell_j^-)=\Gamma(Z\to\ell_i^-\ell_j^+).
\end{equation}
The branching ratio summed over the two charge assignments is
\begin{equation}\label{BRZLCFV}
\text{BR}(Z\to\ell_i^\pm\ell_j^\mp)=\frac{\Gamma(Z\to\ell_i^+\ell_j^-)+\Gamma(Z\to\ell_i^-\ell_j^+)}{\Gamma_Z}=\frac{2\Gamma(Z\to\ell_i^+\ell_j^-)}{\Gamma_Z}.
\end{equation}

\IfFileExists{npb-unsrt.bst}{\bibliographystyle{npb-unsrt}}{\bibliographystyle{unsrt}}
\bibliography{Ref}
\end{document}